\documentclass[preprint2,longabstract]{aastex}

\slugcomment{To be submitted to the Astronomical Journal.}

\shorttitle{The Origin of the 4.5 $\mu$m Excess from Dwarf Galaxies}
\shortauthors{Smith \& Hancock}

\begin{document}


\title{The Origin of the 4.5 $\mu$m Excess from Dwarf Galaxies}


\author{Beverly J. Smith and Mark Hancock\altaffilmark{1}}
\affil{Department of Physics and Astronomy, East Tennessee State University,
    Johnson City TN  37614}

\email{smithbj@etsu.edu}


\altaffiltext{1}{Now at the University of California at Riverside.}


\begin{abstract}

Dwarf galaxies 
tend to have redder [3.6 $\mu$m] $-$ [4.5 $\mu$m] Spitzer broadband
colors than spirals.
To investigate this effect,
for a large sample of dwarf galaxies
we combine
Spitzer fluxes 
with data at other wavelengths and compare to population
synthesis models.
Lower metallicity systems are found to have
redder [3.6] $-$ [4.5] colors on average, but with considerable scatter.
The observed range in [3.6] $-$ [4.5] color is too large to be accounted for solely by
variations
in stellar colors due to age or metallicity differences;
interstellar effects must contribute as well.
For the reddest systems, the 4.5 $\mu$m luminosity may not be
a good tracer of stellar mass.
We identify three factors that redden this color in dwarfs.
First, in some systems, 
strong Br$\alpha$ emission contributes significantly
to the 4.5 $\mu$m emission.
Second, 
in some cases high optical depths lead 
to strong reddening of the starlight 
in the Spitzer bands.   
Third, in some galaxies, 
the nebular continuum dominates the 4.5 $\mu$m
flux, and in extreme cases, the 3.6 $\mu$m flux as well.
The harder UV radiation fields in
lower metallicity systems produce 
both more gaseous continuum in the infrared and 
more Br$\alpha$ per star formation rate.  
The combination of these three factors can account 
for the 4.5 $\mu$m excess in our sample
galaxies, thus it is not necessary to invoke a major
contribution from hot dust to the 4.5 $\mu$m band.
However, given the uncertainties, we are not able to completely rule out
hot dust emission at 4.5 $\mu$m.
More spectroscopic observations in the 3 $-$ 5 $\mu$m range are needed 
to disentangle these effects.

\end{abstract}





\section{Introduction}

The mid-infrared spectra of low metallicity dwarf galaxies
are well-known
to differ from those of spirals, with weaker polycyclic aromatic
hydrocarbon  (PAH) emission
features and higher [Ne~III]/[Ne~II] ratios  \citep{thuan99, madden00,
galliano03,
houck04,
hunt05, 
madden06,
wu06}.
This may be due to harder interstellar radiation fields (ISRFs)
in dwarfs because of lower
dust extinction,
leading to PAH destruction
\citep{galliano03, galliano05,
madden06},
or to delayed creation of carbon dust
by AGB stars in dwarfs
\citep{galliano08}.
These differences are reflected in the mid-infrared broadband
colors of dwarfs, which show depressed 8 $\mu$m emission relative
to 24 $\mu$m, compared to spirals
\citep{boselli98, 
engel05,
rosenberg06, rosenberg08}.
This is
interpreted as a weakening of the 7.7
$\mu$m
PAH feature in dwarfs.
Dwarf and spiral spectral energy distributions also differ at longer
wavelengths;
spiral spectra peak in the far-infrared, while dwarf spectra peak at shorter
wavelengths, $\approx$25 $-$ 60 $\mu$m
\citep{galliano03,
houck04,
hunt05}.
This implies warmer large dust grains in dwarf galaxies on average.
Low metallicity galaxies also tend to have lower total dust-to-gas mass
ratios than higher metallicity systems 
(e.g., \citealp{engelbracht08, hirashita08}).

The Spitzer [3.6] $-$ [4.5] broadband colors of irregular/Sm
galaxies also differ from those of spirals, being redder on average
\citep{pahre04,
engel05, smith07}.
This color difference, however,
has proven more difficult to interpret.
Since the 3 $-$ 5 $\mu$m radiation from galaxies
may be a combination of starlight and interstellar
emission,
it can be difficult to accurately disentangle the contributions
from these components to the observed broadband fluxes.
\citet{pahre04}
suggested that this reddening is due to younger stars
in Irr/Sm galaxies compared to spirals, with hotter stars 
having redder [3.6] $-$ [4.5]
colors. 
In general, hotter stars tend to have redder L $-$ M colors
($\sim$[3.6] $-$ [4.5])
than
cooler stars \citep{cox00}.
Alternatively, these color differences have been attributed
to hot dust in
dwarf galaxies contributing at 4.5 $\mu$m 
\citep{engel05,
hunter06}.
The Br$\alpha$ line may also contribute significantly to
the 4.5 $\mu$m band \citep{churchwell04,
elmegreen06}.
In addition, the nebular continuum in the infrared can
also contribute significantly at wavelengths $\ge$ 3 $\mu$m
in low metallicity systems (e.g., \citealp{kruger95}).
Another factor may be 
the 3.3 $\mu$m PAH feature, which contributes to the 3.6 $\mu$m 
Spitzer band. 
A lower heavy element abundance, and therefore less PAHs, 
can redden the [3.6] $-$ [4.5] color.
Also, stars with low metal abundances may have redder
colors
because of less line blanketing in the 4.5 $\mu$m band
(M. Cohen 2006, private communication).
Evolved stars can also redden
this color,
for example, Miras are
1 $-$ 2 magnitudes redder
in L $-$ M 
than non-variable optically-bright
stars
\citep{smith03}.
In highly obscured systems, reddening due to dust extinction can also be
important even in the Spitzer bands (e.g., \citealp{roussel06}).

To determine the origin of the red [3.6] $-$ [4.5] colors in dwarf galaxies,
for a large sample of
dwarfs 
we have collected Spitzer mid-infrared fluxes,
metallicities,
and ground-based optical, near-infrared, and radio
measurements
from the literature.
Combining these data with
population
synthesis modeling, we have investigated trends in the stellar and interstellar
spectra
with metallicity and age.  
In Section 2 of this paper, we describe the sample and the data.
In Section 3, we compare the [3.6] $-$ [4.5] colors with other properties
of the galaxies.  We discuss expected [3.6] $-$ [4.5] colors from starlight
in Section 4, and in Section 5, we investigate Br$\alpha$ contributions
to the 4.5 $\mu$m band.   In Section 6, we discuss reddening due to
dust extinction in the
Spitzer bands.  In Section 7, we discuss reddening of the [3.6] $-$ [4.5]
color
due to the nebular continuum.
In Section 8, non-stellar contributions to the 3.6 $\mu$m Spitzer band are investigated.
In Section 9, 
the question of hot dust contributions at 4.5 $\mu$m is discussed.
Morphology is discussed in Section 10, while luminosity and stellar mass are
investigated in Section 11.
Some additional issues are discussed in Section 12, and conclusions
are presented in Section 13.

\section{The Sample and the Data}

To investigate how 
the [3.6] $-$ [4.5] color depends upon metallicity and
other properties of 
dwarf galaxies,
we collected Spitzer 3.6, 4.5, 5.8, 8.0, and 24 $\mu$m fluxes 
from the literature
for a sample of 71 
irregular, blue compact dwarf (BCD), Sm, Sd,
and other dwarf galaxies,
for which oxygen abundances are
available.
This sample is not complete in any sense, however,
it spans a large range of metallicity, absolute magnitude, Spitzer
luminosities, and gas content.
Table 1 lists the references 
from which we extracted the data used in this study.

For all of the galaxies in our sample with very red 
or very blue [3.6] $-$ [4.5] published colors, we 
downloaded the Spitzer archival images and checked the published values.
For all except the bluest galaxy
in the sample, UGC 4483, the colors 
we obtained agreed with the published
colors within $\sim$0.15 magnitudes.  
For UGC 4483, however, 
our [3.6] $-$ [4.5]
color is $-$0.16, compared to $-$0.39 from \citet{engelbracht08}.
For this galaxy, we used our fluxes rather
than the published numbers.

We also collected 
total-galaxy
21 cm HI and H$\alpha$ fluxes for the sample galaxies.
We also used
H$\alpha$/H$\beta$ ratios from the literature, preferably
global (total-galaxy) 
values when possible; otherwise, we used smaller aperture measurements.
In addition, we searched for 
both total-galaxy and smaller aperture Br$\alpha$ and Br$\gamma$ measurements.
We also obtained 
H$\beta$ equivalent widths and measurements of the 9.7 $\mu$m
silicate absorption feature optical depth for our sample galaxies, preferring
global values over small aperture measurements when available.
In Table 1, for each type of data,
we give the number of galaxies with such data available.

We also obtained  
total-galaxy 
optical UBVRI and 
near-infrared JHK broadband fluxes for our 
sample from the literature.
These magnitudes were corrected for 
Galactic extinction, when necessary, and 
converted to Jy using the zero points in \citet{bessel98}.
In a number of cases, 
large discrepancies were found between different published 
magnitudes
for the same galaxy.
In these cases, we did not use either measurement in our analysis.

As a comparison sample of spiral galaxies, we started with the sample of nearby `normal'
(not strongly interacting) spirals discussed in \citet{smith07}.
These galaxies have total-galaxy Spitzer infrared and ground-based
optical and near-infrared broadband fluxes available from
\citet{dale07}.   Most have total H$\alpha$ fluxes available from
\citet{kennicutt03}, and Pa$\alpha$/H$\alpha$ ratios for the inner
50$''$ available from \citet{calzetti07}.
As spirals tend to have radial abundance gradients,
for comparison to the dwarfs we selected spirals which have disk-averaged
oxygen
abundances available \citep{mous06, calzetti07}.
After removing Sd galaxies, 
there are 13 galaxies in the spiral sample.

\section{[3.6] $-$ [4.5] Colors 
as a Function of Metallicity, Gas Content, and Star Formation Rate }

In the left panel of Figure 1,
we compare the [3.6] $-$ [4.5] color against oxygen abundance
for our sample dwarf galaxies.
The right panel shows the same plot for the spirals.
The dwarfs are on average redder in [3.6] $-$ [4.5], in addition to
having lower metallicities.
For the dwarfs, a relation is seen, in that
colors redder than [3.6] $-$ [4.5] $>$ 0.3 are only seen at
low abundances,
log(O/H) + 12 $\le$ 8.2 ($<$ 0.5 solar).
In contrast, the spirals all have [3.6] $-$ [4.5] $\sim$ 0.0, and
disk-averaged log(O/H) + 12 $>$ 8.2.

For the dwarfs, there is considerable scatter in Figure 1.
To test whether this scatter is due solely to differing amounts of
interstellar matter,
with more gas-rich galaxies having a larger amount of dust and therefore
more 4.5 $\mu$m excess,
in Figure 1 we color-coded the data points according to their 
m(HI) $-$ [3.6] color, 
where m(HI) is the magnitude
in the 21 cm HI line \citep{devauc91, paturel03}.
The corresponding HI mass for the dwarfs range from
7 $\times$ 10$^6$ M$_{\sun}$ to 7 $\times$ 10$^9$ M$_{\sun}$.
In this plot,
as log(O/H) increases, m(HI) $-$ [3.6]
tends to 
increase 
(i.e., the HI mass to 3.6 $\mu$m luminosity ratio M$_{HI}$/L$_{3.6}$ decreases).
This is consistent with results of 
previous studies (e.g.,
\citealp{fisher75}),
which show that 
lower metallicity systems tend to be richer in HI relative to
their stellar mass. 
However, 
for a given metallicity, no strong
trend of 
m(HI) $-$ [3.6]
with [3.6] $-$ [4.5] is found.
Thus the 4.5 $\mu$m excess does not appear to be due to an increasing mass
fraction of
interstellar matter.
This result is uncertain, however, because molecular gas is not included in 
this plot.
Many of these galaxies are undetected in CO, the standard
tracer of molecular gas in galaxies, and the amount of molecular
gas present is uncertain.

In Figure 2, we compare the [3.6] $-$ [4.5] colors of these
galaxies with
the
[3.6] $-$ [24] colors.
The left panel shows the dwarfs, the right the spirals.
In this Figure,
we have color-coded the data points as a function of oxygen abundance.
The 24 $\mu$m luminosity is considered a good tracer
of star formation rate, as it arises from small dust
grains 
heated by UV photons
(e.g., \citealp{calzetti05}).
If the 3.6 $\mu$m flux is mainly due to older stars, then 
[3.6] $-$ [24] is roughly a measure of the mass-normalized star
formation
rate, and therefore the intensity of the UV field.
Figure 2 shows that indeed more intense UV fields are 
correlated with more 4.5 $\mu$m
excess.
The distributions of the spirals and the dwarfs overlap in Figure 2, but
the dwarf sample extends to much redder [3.6] $-$ [24] colors as well
as redder [3.6] $-$ [4.5] colors than the spirals.
The dwarf galaxies with the redder colors tend to have lower metallicities,
however, some low metallicity galaxies have blue Spitzer colors.
Thus low metallicity appears to be a necessary but not sufficient
condition for red [3.6] $-$ [4.5] and [3.6] $-$ [24] colors.

\section{Stellar [3.6] $-$ [4.5] Colors from Population Synthesis }

To test whether stars alone can account for the variations in
[3.6] $-$ [4.5] colors, we 
used the 
Starburst99 stellar population synthesis code
\citep{leitherer99}
to produce model stellar
colors in these bands.   We used version 5.1 of this code,
which includes the Padova asymptotic giant branch stellar evolution
models \citep{vazquez05}.
We note that 
this code does not 
model 
red supergiants in low metallicity
systems well \citep{vazquez05},
which introduces some uncertainty to our results.
We used a grid of models 
with ages that
range from 1 Myrs to 20 Gyrs, 
and metallicities of 1/50, 1/5, 1/2.5, 1, and 2.5 times solar.  

Dwarf galaxies have a large range of mass-normalized star formation rates;
some are undergoing violent bursts of star formation, while others have low
current rates \citep{hunter97}.
The integrated UBV colors of dwarf irregular galaxies are generally consistent
with
constant star formation rates 
\citep{hunter85,
vanZee01}.
Blue compact dwarf galaxies 
are undergoing recent starbursts, but
an older population
is usually also present
\citep{shulte98, shulte01,
droz02,
kong03,
thuan05}.
As limiting cases to the true star formation histories of the galaxies,
we therefore ran
both continuous and instantaneous burst star formation models.

We integrated the Starburst99 model spectra over the 
Spitzer bandpasses and calculated the [3.6] $-$ [4.5] colors.
In Figure 3, we plot model [3.6] $-$ [4.5] colors vs.\
age for various metallicities.
In Figure 3, we only include stellar contributions to
the Spitzer fluxes; no dust emission or interstellar gas emission lines are
included.   
The models shown in Figure 3 also do not have reddening due to dust extinction
included (see Section 6).
Color corrections are also not included, as these
are expected to be very small (see IRAC Data Manual).  
In the top panel,
we present instantaneous burst models, 
while the bottom panel
shows continuous star formation models.

Figure 3 shows that, according to Starburst99,
there is a relatively small spread in the [3.6] $-$ [4.5] colors
of stars of different ages and metallicities, $\le$0.2 magnitudes.
As expected, there is 
a trend such that younger ($\le$6 $\times$ 10$^6$ Myrs)
and lower metallicity systems (except $\ge$2.5 Z$_{\sun}$)
have redder colors than intermediate
age systems ($\le$10$^8$ Myrs).  However,
the spread in colors is quite small, and cannot account for the full range of
observed colors for dwarfs.
Except for very young ($\le$5 Myrs) models with greater than solar
metallicity, the model stellar 
colors are between $-$0.16 and 0.04.
Of our dwarf galaxies, 70\% are redder
than this limit.
Thus the observed range in [3.6] $-$ [4.5] colors of dwarfs is not due 
solely to variations in star colors due to age and metallicity variations.

\section{Br$\alpha$ Contributions to the 4.5 $\mu$m Spitzer Band }

In this section, we estimate the importance of the 
Br$\alpha$ line at 4.05 $\mu$m, which may contribute
significantly to the Spitzer 4.5 $\mu$m 
band (e.g., \citealp{churchwell04}).   
For example, in the Galactic H~II region M17, Br$\alpha$
is estimated to contribute 20\% of the total 4.5 $\mu$m broadband flux
\citep{povich07}.
At low metallicities, 
stars are hotter and therefore more ionizing photons are available
for the same mass of stars.  Thus as metallicity decreases,
for the same star formation rate the 
H$\alpha$ and 
Br$\alpha$ fluxes are expected to increase (e.g., \citealp{charlot01, lee02}).

To determine how much Br$\alpha$ can affect the [3.6] $-$ [4.5] color, 
we added Br$\alpha$ to the Starburst99 models shown in Figure 3.
We used the H$\alpha$ fluxes provided by Starburst99,
along with a nominal Br$\alpha$/H$\alpha$ ratio of 0.0273 (10,000 K,
Case B; \citealp{hummer87}).
We then added this model Br$\alpha$ flux to the Starburst99 stellar spectrum
and integrated over the Spitzer bandpasses to get model colors.
In Figure 4, we plot the [3.6] $-$ [4.5] model color including 
Br$\alpha$ against age, with different metallicity models
plotted with different symbols.   These models do not include
dust extinction or emission, or other interstellar lines besides
Br$\alpha$.   
A comparison of Figure 4 to
Figure 3 shows that, in some very young systems,
Br$\alpha$ can
contribute significantly,
particularly for
low metallicity systems.   
For example, for ages $\le$ 10 Myrs and a metallicity of 1/50 solar,
Br$\alpha$ may redden the [3.6] $-$ [4.5] color by $\ge$1 magnitude.
A higher assumed temperature will decrease this reddening somewhat.

A more direct way to determine the Br$\alpha$ contribution to
the observed 4.5 $\mu$m flux is to either measure it directly,
or to extrapolate from Br$\gamma$ or H$\alpha$ measurements.
Only one galaxy in our sample, 
II Zw 40, had a total-galaxy Br$\alpha$ flux available
\citep{verma03}.
From this measurement, 
we find that only $\sim$11\%
of the Spitzer 4.5 $\mu$m flux of II Zw 40
is due to Br$\alpha$.
Another galaxy in our sample
has a total-galaxy published Br$\gamma$ flux available 
(SBSG 0335-052; \citealp{dale01}). Using the small-aperture spectroscopic
Br$\alpha$/Br$\gamma$ flux ratio from \citet{vanzi00}
and \citet{hunt01} to extrapolate to
the total Br$\alpha$ flux, we estimate that the Br$\alpha$
contribution to the 4.5 $\mu$m Spitzer flux for this galaxy is
$\sim$31\%.   This may be an over-estimate, since some of the Br$\gamma$
flux may arise in less extincted regions (e.g., \citealp{vanzi00}).

For NGC 5253, the Br$\alpha$ measurement of 
\citet{kawara89} in a 10$''$ $\times$ 20$''$ aperture
gives a lower limit to the Br$\alpha$ contribution to the Spitzer 4.5 $\mu$m flux of 2$\%$.
This aperture is
considerably smaller than
the optical angular size for NGC 5253 
of $\sim$2$'$ $\times$ 5$'$
given in the NASA Extragalactic Database (NED) 
and the observable extent of the 3.6 $\mu$m emission in the archival Spitzer image of $\sim$1$'$ $\times$ 2$'$.
However, the infrared emission is strongly centrally peaked in the archival Spitzer
images, with 25$\%$ of the total
3.6 $\mu$m flux 
and 47$\%$ of the 4.5 $\mu$m flux within a 10$''$ $\times$ 10$''$ region.
Thus the Br$\alpha$ contribution to the 4.5 $\mu$m
flux is likely not much more than 4 $-$ 8$\%$ for NGC 5253.

For galaxies with published total-galaxy H$\alpha$ measurements, we estimate the 
Br$\alpha$ contribution indirectly from these fluxes,
correcting for 
internal extinction
using the published H$\alpha$/H$\beta$ ratio.
From the extinction-corrected H$\alpha$ flux, we calculated the
Br$\alpha$ flux, and determined the fraction of the observed 4.5 $\mu$m
broadband flux due to Br$\alpha$.  
This calculation shows that, if the H$\alpha$/H$\beta$ estimates
of extinction are appropriate,
for our sample galaxies
the fraction of the observed Spitzer flux due to 
Br$\alpha$
ranges from 0.3\% to 23\%,
with most (80\%) having less than 10\% of their total 4.5 $\mu$m flux
due to Br$\alpha$.   
In some cases, H$\alpha$/H$\beta$ may underestimate the extinction (see Section 6),
thus these percentages are lower limits.

We then re-calculated the [3.6] $-$ [4.5] colors of the galaxies, after
removing the flux due to Br$\alpha$.
In the left panel in Figure 5, for the dwarf galaxies with published H$\alpha$ fluxes,
we plot the metallicity against both the observed
and the Br$\alpha$-removed
[3.6] $-$ [4.5] colors.
For the galaxies with measured Br$\alpha$ or Br$\gamma$, 
we plot two Br$\alpha$-corrected
values, one calculated using the H$\alpha$ (filled red squares) and one 
using the Br$\alpha$ or Br$\gamma$
value (open green triangles).
In Figure 5, we mark the range in [3.6] $-$ [4.5] expected
from starlight alone,
$-$0.16 $\le$ [3.6] $-$ [4.5] $\le$ 0.04 (see Figure 3).
The correction for Br$\alpha$ shifts the galaxies to the left,
to bluer [3.6] $-$ [4.5] colors, with the shifts ranging from 0 $-$ 0.3 
magnitudes.  
Note that for II Zw 40 and SBSG 0335$-$052, the Br$\alpha$ correction inferred
from H$\alpha$ and H$\alpha$/H$\beta$ measurements is less than that 
determined from a direct Br$\alpha$ measurement (in the case of II Zw 40)
or from Br$\gamma$ (for SBSG 0335$-$052).   This suggests that
the H$\alpha$/H$\beta$ ratio may underestimate the extinction
in some cases (see Section 6).

Most of the
shifts seen in Figure 5 are relatively small, $\le$0.1 magnitudes.
However, for a few of our galaxies (for example, He 2-10), 
the shifts are sufficient to move
the galaxies into the range expected from starlight.
In these cases, Br$\alpha$ may be mainly responsible for the red
[3.6] $-$ [4.5] colors.

However, for some of our sample galaxies (e.g., II Zw 40,
SBSG 0335-052, 
HS 0822+3542, 
NGC 4194, and SHOC 391),
the Br$\alpha$-corrected colors are still redder than
those expected by starlight alone.
For II Zw 40, this conclusion is particularly robust, because
a direct Br$\alpha$ measurement is available \citep{verma03}.
For SBSG 0335-052, the Br$\gamma$ observation
of \citet{dale01} also provides a reasonably strong constraint.

In the right panel of Figure 5, a similar plot is shown for the spiral
sample.  The open black circles are the published [3.6] $-$ [4.5] values, while the 
red crosses show the values after correction for Br$\alpha$, calculated
using the total-galaxy H$\alpha$ fluxes and the Pa$\alpha$/H$\alpha$ ratio
for the inner 50$''$ \citep{calzetti07}.
This plot shows that for normal spirals the 
corrections for Br$\alpha$ in the Spitzer
4.5 $\mu$m filter is generally very small,
consistent with the fact that most of the published [3.6] $-$ [4.5]
values are in the expected range for starlight.

\section{Reddening of Starlight in the Spitzer Bands due to Dust Extinction}

Another possible source of reddening of the [3.6] $-$ [4.5] color is dust 
extinction of the starlight.
The model [3.6] $-$ [4.5] colors of stars shown in Figure 3
do not include reddening due to dust extinction.  
In modeling the spectral energy distributions
in galaxies, reddening is sometimes neglected
at wavelengths $\ge$ 3.6 $\mu$m (e.g., \citealp{draine07}).
However, if the infrared sources are highly obscured, reddening by extinction
may be important, even in the Spitzer bands.

How much reddening is present depends upon the properties
of the dust grains, and therefore the extinction law, as well as
the geometry of the system.  The ratio of visual extinction
to B $-$ V reddening A$_V$/E(B $-$ V) = R$_V$ 
has been found to vary from $\sim$3.1
in diffuse gas to $\sim$5.5 in dense clouds (e.g., \citealp{cardelli89}).
The extinction law in the Spitzer bands also varies with optical
depth \citep{chapman08}.  Combining the extinction laws from
these papers gives
a range for A$_V$/E(3.6 $-$ 4.5) between $\sim$ 41 for diffuse gas
to $\sim$68 for more extincted regions.

The published H$\alpha$/H$\beta$ ratios for our sample dwarfs
imply A$_V$ between 0 and 2.5, or 
reddening in [3.6] $-$ [4.5] of less
than 0.06 magnitudes, not sufficient to account for the 4.5 $\mu$m
excess in our reddest systems.
However, there is evidence that the H$\alpha$/H$\beta$
ratio strongly underestimates the extinction in at least some of our systems.

The H$\alpha$/H$\beta$ ratio of SBSG 0335-052, 
the galaxy in our sample with the reddest [3.6] $-$ [4.5] color, 
implies only A$_V$ $\sim$ 0.3 $-$ 0.5 
\citep{izotov97, izotov07}.
However, 
mid-infrared spectroscopy 
of 
SBSG
0335-052 
shows
a deep 9.7 $\mu$m silicate absorption feature 
\citep{thuan99,
houck04}, suggesting a much
higher optical depth of A$_V$ $\sim$ 15 $-$ 20.
The Br$\gamma$/Br$\alpha$ ratio also suggests a much larger extinction
than that implied by
the H$\alpha$/H$\beta$ ratio, A$_V$ $\sim$ 12 \citep{hunt01}.
Dust modeling of the 6 $-$ 100 $\mu$m spectral energy distribution
also 
suggests that most of the mid-infrared emission originates
from a highly obscured (A$_V$ $\sim$ 12$-$30) luminous star 
cluster \citep{hunt01, hunt05, plante02,
takeuchi03}.
The increase in derived extinction with wavelength
implies a more complex geometry than a simple
uniform foreground screen of dust, and suggests
that the light seen in the optical
is less obscured than the bulk of the starburst
(e.g., \citealp{hunt01, reines08}).
The 3$-$4 $\mu$m source 
in
SBSG
0335-052 
is very compact, $\le$1.2$''$ \citep{hunt01},
consistent with this picture.
If the extinction obtained from the silicate absorption,
A$_V$ $\sim$ 15$-$20,
applies
to the majority of the stars contributing to the 3.6 and 4.5 $\mu$m emission,
then E(3.6 $-$ 4.5) $\sim$ 0.22$-$0.48.  Combined with Br$\alpha$, this
can account for some, but not all, of the observed reddening in
[3.6] $-$ [4.5].

In the left panel of Figure 6, 
we plot the oxygen abundance 
of the sample dwarf galaxies
against the [3.6] $-$ [4.5] color, after this color has been
corrected for both Br$\alpha$ and reddening of the
starlight due to dust extinction.
For these galaxies,
we plot extinction corrections made using 
the 9.7 $\mu$m silicate optical depth (blue crosses), 
the 
Br$\alpha$/Br$\gamma$ ratio (magenta asterisks), and/or the H$\alpha$/H$\beta$ ratio (black
filled diamonds), 
depending upon the available data.
As in Figure 5, we have marked the expected colors due to starlight alone.
Note that the [3.6] $-$ [4.5] color of
SBSG 0335-052 is too red to be accounted for by stars, even after correcting
for both Br$\alpha$ and extinction.  Thus there must be another contributor to the 4.5 $\mu$m
excess for this galaxy.

A similar situation may exist for 
II Zw 40. 
The H$\alpha$/H$\beta$ ratio 
implies A$_V$ $\sim$ 1.1
\citep{french80,
kinman81},
or E(3.6 $-$ 4.5) $\sim$ 0.03.
However, longer wavelength data gives higher extinctions.
The Br$\alpha$/Br$\beta$ ratio gives A$_V$ $\sim$ 10 \citep{verma03},
while comparison of Br$\gamma$ to radio continuum implies A$_V$ $\sim$ 
8 $-$ 10
and modeling of the infrared spectral energy distribution
gives A$_V$ $\sim$ 20 $-$ 30 \citep{hunt05}.
As with 
SBSG
0335-052, 
the mid-infrared source in II Zw 40 is very compact,
with most emission confined to 0.5$''$ (Beck et al.\ 2002).
In contrast to
SBSG
0335-052, 
however,
no silicate absorption feature is seen
in mid-infrared spectra of II Zw 40
\citep{martin06,
wu06},
thus the extinction is uncertain.
If A$_V$ $\sim$ 10 holds for the bulk of the population contributing
to the Spitzer broadband flux, then E(3.6 $-$ 4.5) $\sim$ 0.15 to 0.24.  
As shown in Figure 6, this could account for some, but not all, of the observed
reddening. 

NGC 5253 is another example of a galaxy with a larger implied
optical depth from the Br$\alpha$/Br$\gamma$ ratio (A$_V$ $\sim$ 11 in a 10$''$ $\times$
20$''$ beam; \citealp{kawara89}) and the silicate absorption (A$_V$ $\sim$ 8 $-$ 25 in
a 5\farcs4 aperture; \citealp{aitken82}) than from the H$\alpha$/H$\beta$ ratio
(A$_V$ $\sim$ 0.3 for the whole galaxy; \citealp{mous06}).
As with II Zw 40 and 
SBSG
0335-052, extinction and Br$\alpha$ together cannot account for all of the reddening
in [3.6] $-$ [4.5] (see Figure 6).  
As noted in Section 5, however, our correction for Br$\alpha$ may be slightly
too small.

For He 2-10 (ESO 495-21), 
if the silicate absorption (A$_V$ $\sim$ 15 in a 5\farcs9 aperture; \citealp{phillips84})
or the Br$\alpha$/Br$\gamma$-implied extinction (A$_V$ $\sim$ 17 in 7\farcs1 $\times$ 3\farcs5 aperture;
\citealp{kawara89}) are used to correct the Spitzer colors, instead of the extinction
implied by the H$\alpha$/H$\beta$ ratio (A$_V$ $\sim$ 1.7; \citealp{vacca92}), the corrected
[3.6] $-$ [4.5] color becomes too blue (see Figure 6).  In this galaxy,
the lower extinction estimate may be more appropriate.

Two of our other dwarfs with red [3.6] $-$ [4.5] colors,
NGC 4194 and Haro 11 (ESO 350-38), also show apparent silicate absorption
in their mid-infrared spectra \citep{brandl06,
wu06}.
For Haro 11, no silicate optical depth analysis has been done yet.
For NGC 4194, the silicate optical depth is estimated to
be $\tau$$_{9.7}$ = 0.37 \citep{brandl06}.  Assuming
A$_V$/A$_{9.7}$ $\sim$ 18.5 \citep{roche84} gives
E(3.6 $-$ 4.5) between 0.10 and 0.17.  Combined with the correction for
Br$\alpha$, this can account for most,
but not all, of the 
excess 4.5 $\mu$m emission (see Figure 6).

In contrast to these
galaxies, I Zw 18 
shows no silicate absorption \citep{wu07} and the extinction
implied by the H$\alpha$/H$\beta$ ratio is very small
\citep{cannon02}.
In this case, the correction
for
extinction is small compared to that for 
Br$\alpha$ 
(see Figures 5 and 6).   

One caveat in this analysis is that
the reddening of the stars in a galaxy
may differ significantly from that of the ionized gas,
depending upon
the geometry of the system.
Thus
caution should be taken in applying 
extinction corrections obtained
from emission line ratios or silicate absorption features to stellar colors.   
Furthermore, for different galaxies different datasets are
available.

In the right panel of Figure 6, 
the blue open diamonds show the [3.6] $-$ [4.5] colors
of the normal spirals 
after correction for both Br$\alpha$
and for starlight reddening.  The latter term
was calculated using the Pa$\alpha$/H$\alpha$
ratios of \citet{calzetti07}.   Note that these corrections are
small for the spirals, with A$_V$ ranging
from 0.2 $-$ 2.5.  
Thus the correction for starlight
reddening, like that for Br$\alpha$, does not shift
the spirals much on this plot.
We note that other techniques for obtaining extinctions such
as background counts of galaxies
through galaxian disks also give relatively small average obscurations
for nearby spirals \citep{white00}, 
consistent with the above 
range in A$_V$ for spirals (see also
\citealp{calzetti01}).

\section{Contributions from the Nebular Continuum }

Another source of near- to mid-infrared light in galaxies is nebular
continuum emission, which we have neglected so far in our analysis.
In the case of low metallicity starburst systems, this component
can be important \citep{joy88,
kruger95,
vanzi00,
hunt01}.
To determine the contribution from the gaseous continuum
to the Spitzer bands, we ran another set of Starburst99 models, including
just starlight and the gaseous nebular continuum, excluding
Br$\alpha$ and dust reddening.  
To calculate the gaseous continuum, Starburst99 uses
parameters from \citet{ferland80}. The longest wavelengths used in
these tables is 4.5 $\mu$m, therefore the predicted values of the nebular
continuum are unreliable at longer
wavelengths \citep{leitherer99}.

In Figure 7, we plot
the model [3.6] $-$ [4.5] colors as a function of age for
various metallicities.
Comparison with Figure 3 shows that
the addition of nebular continuum emission significantly reddens the
[3.6] $-$ [4.5] colors of young systems, particular for low metallicities.
For example, $\le$10 Myr 0.02 Z$_{\sun}$ instantaneous burst models
with the nebular continuum included
have 
[3.6] $-$ [4.5] colors about 0.5 magnitudes redder than models
without this component.
For solar metallicities, 
ages $\le$4 Myrs produce such red colors.
Thus the nebular continuum may be a very important component in
very young and/or low metallicity systems.

In extreme cases, the nebular continuum may dominate not only the 4.5 $\mu$m
band, but also the 3.6 $\mu$m band.  This is 
illustrated in Figures 8 and 9, where we 
plot the ratio of the nebular continuum flux 
in the 3.6 $\mu$m and 4.5 $\mu$m bands
to the flux from starlight plus the nebular continuum,
for the models shown in Figure 7.
Figures 8 and 9 show that, for solar metallicity and ages $\le$ 3 Myrs,
and for 0.02Z$_{\sun}$ and $\le$ 10 Myrs, contributions to both
bands from the
nebular continuum are larger than from starlight.   
\citet{kruger95}
found that, for a 5 Myr 0.02Z$_{\sun}$ burst, 
$\sim$70$\%$ of the flux in the L band is due to the nebular continuum,
consistent with these plots.
Since the nebular contributions are typically larger at 4.5 $\mu$m
than 3.6 $\mu$m, the net effect is a reddening of the observed
colors of galaxies.
The issue of non-stellar emissions at 3.6 $\mu$m is discussed
further in Section 8.

To
determine the contribution from the nebular continuum for individual
galaxies in our sample, we 
require an estimate of the average age of the stellar population.
We do this two ways.   First, we use Starburst99 models
and published H$\beta$ equivalent widths to estimate the age.
For each galaxy,
we use models with a metallicity closest to that of the
published
metallicity, and we fit
both instantaneous
and continuous star formation models.
Second,
when sufficient broadband optical 
data is available ($\ge$3 filters), we fit
the optical spectral energy distribution
to 
Starburst99 models to get
the age and extinction of the 
contributing stars, and the uncertainties on these values (e.g., \citealp{smith08,
hancock08}).
We include the nebular continuum in the optical
in doing these fits, and we conservatively assumed uncertainties in the optical
colors of 0.1 magnitudes for all galaxies.
In convolving the model spectra with the response functions of the broadband filters,
we added the H$\alpha$ emission line (e.g., \citealp{smith08}),
but did not include other emission lines.
We used the \citet{cardelli89} extinction law.

We did not use the near-infrared
photometry in the fits, but used
them after the fact to decide
whether instantaneous or continuous star formation is more appropriate
for the system.                                                                    
In most cases, both instantaneous and continuous models were consistent with
the near-infrared data; in a few galaxies, the continuous models
produced too much near-infrared or 3.6 $\mu$m flux.

We then used the best-fit ages from the instantaneous burst models
to determine the amount of
reddening in the [3.6] $-$ [4.5] color 
due to the nebular continuum.
We then corrected the observed colors by this amount, adding to the
corrections for 
Br$\alpha$ and dust extinction discussed above.
As discussed in Section 4, some dwarf galaxies can be fit by
continuous star formation models, while others are better explained
by multiple bursts.
For normal spirals, the broadband optical spectral energy distributions
can 
generally be fit by continuous or exponentially decaying star formation
\citep{searle73, larson78}.
We choose to use instantaneous burst models to calculate the
nebular contribution to the 4.5 $\mu$m band since
they produce less nebular 
emission for the same age, as shown in Figures 8 and 9.
This thus provides a limiting case for the nebular continuum
contribution.
For some galaxies, the fit to the optical data was quite
poor, likely because their star formation history
does not match either an instantaneous or continuous burst.
These galaxies were excluded from Figure 10.

In Figure 10,
the nebular-corrected [3.6] $-$ [4.5]
colors are plotted against oxygen abundance,
with the dwarfs in the left panel and the spirals in the right.
In these plots, the filled upside-down magenta triangles 
are the nebular continuum-corrected colors calculated using the best-fit ages from the H$\beta$
equivalent widths, while the cyan open diamonds
are the nebular-corrected values using the best-fit ages from the broadband
optical fitting.
To illustrate the uncertainties in this correction, 
the open red circle
shows the color after correction for nebular emission using the best fit age 
from the broadband fitting
minus the 1$\sigma$ uncertainty
in the age. 
In some cases, the corrections to the Spitzer [3.6] $-$ [4.5] color due to the nebular
continuum is quite large, for example, it is $\sim$0.4 magnitudes for SBSG
0335-052.

In most cases, within the uncertainties
the combined contributions of Br$\alpha$, reddening due to dust
extinction, and the nebular continuum
can account for all of the observed reddening above starlight.
In Figure 10, for all of the galaxies except two, 
the two estimates of the corrected
color (the magenta upside-down filled triangle and the 
cyan open diamond) or the lower limit on the corrected color (open red circle)
lie within the green hatched region, or span this region.
For the two exceptions, 
the datasets are incomplete, thus we cannot make firm conclusions.
For Haro 11, as noted in Section 6, silicate absorption may be present in the Spitzer spectrum 
\citep{wu06}, however, no 
optical depth has been published, and no published Br$\alpha$ or Br$\gamma$
values are available.   
For SHOC 391, no Br$\alpha$, Br$\gamma$, or silicate absorption measurements
are available.
Thus the extinction of the starlight and/or the correction for Br$\alpha$
in the Spitzer 4.5 $\mu$m band may be under-estimated for these galaxies.

\section{Non-Stellar Contributions to the 3.6 $\mu$m Band }

It is often assumed that the 3.6 $\mu$m band is dominated
by starlight, and a standard [3.6] $-$ [4.5] spectrum
is assumed for the stellar component, which is
then
subtracted from the observed 4.5 $\mu$m flux to
get the residual 4.5 $\mu$m emission 
(e.g., \citealp{helou04,
engel05,
dale05}).
However, as noted above, nebular continuum emission
may be important at both 3.6 $\mu$m and 4.5 $\mu$m (see Section 7).   Also,
if hot dust is significant at 4.5 $\mu$m, it may also be 
present at 3.6 $\mu$m, and even at 2 $\mu$m
(e.g., \citealp{lu03}).

To test whether significant non-stellar emission 
is present 
at
3.6 $\mu$m 
in individual galaxies in our sample, we used our Starburst99 fits to the
broadband optical data.
For all but one of the galaxies with available optical data,
we cannot rule out that all of the 3.6 $\mu$m emission is due to 
starlight, since our best-fit stellar spectra are consistent with
the observed Spitzer 3.6 $\mu$m flux within the uncertainties.
The only galaxy in our sample that shows a clear 3.6 $\mu$m
excess above the inferred stellar continuum
is the reddest galaxy in [3.6] $-$ [4.5],
SBSG 0335-052 
(see Figure 11; also see \citealp{galliano08}).
For SBSG 0335-052,
although
the inferred stellar continuum 
is consistent with the near-infrared photometry for both the instantaneous
and continuous bursts, it 
is about a factor of 5 below
the Spitzer 3.6 $\mu$m measurement.
However, when the nebular continuum is included in the Spitzer bands,
most of the 3.6 $\mu$m 
flux can be accounted for
(see Figure 11).
This shows that the 3.6 $\mu$m excess may be mainly due to the 
gaseous continuum.

We note that the best-fit extinction for the Starburst99 model
of SBSG 0335-052, A$_V$ = 1.0 $\pm$ 0.5,
is significantly less than that inferred by the 
9.7 $\mu$m silicate absorption feature (A$_V$ $\sim$ 15 $-$ 20; \citealp{thuan99,
houck04}).   As mentioned in Section 6,
for this galaxy, the stars that dominate the observed light in the optical 
may be much less obscured
than the stars that dominate in the infrared.   Thus our Starburst99-derived
extinction based on the optical data may not be an appropriate value to use in the infrared.
A second more highly obscured stellar population, 
unseen in the optical,
may also 
contribute to the observed Spitzer 3.6 $\mu$m flux of this system.
However, this second population is not evident in the near-infrared data.

Unfortunately, for some galaxies in our sample, no suitable broadband
optical data were available for fitting, or we were not able to
get good fits to the available
data with our Starburst99 models.  Thus we were not able to
use this method to test for 3.6 $\mu$m excesses above the stellar continuum
in these galaxies.   These galaxies include some of the reddest
galaxies in our sample in [3.6] $-$ [4.5]: SHOC 391, Haro 11, Pox 4, and
HS 0822+3542.   Further observations and more detailed modeling of these
systems are needed to search for 3.6 $\mu$m excesses above the stellar
continuum.

None of the spirals in our sample for which a good fit to
the optical data was found showed a 3.6 $\mu$m excess above
the inferred stellar continuum.  As they also have [3.6] $-$ [4.5]
colors consistent with starlight (Figure 1), this also implies little
4.5 $\mu$m non-stellar excess in the spirals.   
As noted earlier, the 3.3 $\mu$m
PAH feature lies within the 3.6 $\mu$m Spitzer band,
however, the spirals in our sample do not show unusually
blue [3.6] $-$ [4.5] colors. Thus there is no strong evidence
for a strong PAH contribution to the 3.6 $\mu$m flux
of the spirals in our sample.
As noted earlier, dwarf galaxies tend to have weaker PAH
features than spirals, thus this conclusion also may
hold for the dwarf galaxies.

An alternative way to search for 3.6 $\mu$m 
excesses is to compare the [3.6] $-$
[4.5] colors with other near-to-mid-infrared colors (Figure 12).
The spirals (right panel) show less scatter in these colors
than the dwarfs (left panel), and the higher metallicity dwarfs
(blue open squares) show less scatter than the lower metallicity
dwarfs.
For the dwarfs,
weak correlations
are seen between [3.6] $-$ [4.5] and K $-$ [3.6],
and between [3.6] $-$ [4.5] and 
H $-$ [3.6], in that the
extremely red galaxies in [3.6] $-$ [4.5] also tend to be red in 
these other colors.   This suggests that non-stellar contributions
at 3.6 $\mu$m 
may be present in these systems.
The correlation seen between [3.6] $-$ [4.5] and K $-$ [4.5] is
not unexpected, if K roughly traces the older stellar
population.
No correlation is seen between [3.6] $-$ [4.5] with H $-$ K.

\section{Contributions at 4.5 $\mu$m from Hot Interstellar Dust }

It is often assumed that excess emission at 4.5 $\mu$m
is due to hot dust
(e.g., \citealp{engel05, hunter06}).
However, Figure 10 shows that, for all the galaxies in our
sample with sufficient data available, a combination of Br$\alpha$, the nebular continuum,
and dust reddening may be able to
account for most of the 4.5 $\mu$m excess.  Thus a major
contribution from dust emission is not required.  However,
with the available data, given the uncertainties we are not able to completely rule
out hot dust emission in the 4.5 $\mu$m band.

One way to distinguish between these different
processes is with spectroscopy.  However,
at the present time, few high S/N spectra 
of galaxies
in the 3 $-$ 5 $\mu$m range are available, since the Spitzer
spectrometers do not reach wavelengths $\le$ 4.9 $\mu$m.
Co-added Infrared Space Observatory (ISO) 2.5 $-$ 4.9 $\mu$m spectra
of disk galaxies show a continuum excess with a color temperature
of $\sim$1000K \citep{lu03}.
At longer wavelengths, in the $\sim$24 $\mu$m range, 
the continuum emission from star forming galaxies is 
attributed to 
`very small grains' (VSGs; \citealp{desert90}).  
In Galactic HII regions, the 6 $-$ 16 $\mu$m 
VSG continuum is strong, 
in contrast to the PAH features which weaken in HII regions
\citep{cesarsky96,
peeters04}.
Within dwarf galaxies,
the ratio of the PAH feature strength to the VSG continuum
varies spatially, being anti-correlated
with the
[Ne~III 15.8 $\mu$m]/[Ne~II 12.8 $\mu$m] line ratio
\citep{madden06, beirao06}.
Since 
[Ne~III]/[Ne~II] is a measure of radiation field
hardness, and 
metallicity is not expected to vary
much within a dwarf galaxy, this indicates
that PAHs are destroyed in hard UV fields,
while mid-infrared VSG excitation is enhanced.

In a few 
low metallicity galaxies 
with
spatially-resolved 
ISO spectra, 
the short wavelength end (4.9 $-$ 5.6 $\mu$m)
shows strong continuum emission in regions with
strong H$\alpha$ emission 
\citep{madden06}.
This supports the idea
that the VSG continuum may extend well into the 3 $-$ 5 $\mu$m range.
For 
SBSG 0335-052, 
a Spitzer spectrum 
shows a 5 $\mu$m continuum $\sim$ 2 mJy \citep{houck04}.
Extrapolation of this to shorter wavelengths could possibly account 
for much of the observed 4.5 $\mu$m broadband emission.
Three of our galaxies with very red [3.6] $-$ [4.5] colors,
SBSG 0335-052, Haro 11, and SHOC 391, have been 
classified as `mid-infrared peakers' by \citet{engelbracht08}, 
having infrared spectral energy distributions that peak near 24 $\mu$m
rather than in the far-infrared.   This is consistent with these
galaxies having very hot dust grains on average, and the
dust continuum extending down to $\le$4.5 $\mu$m.  

Thus with the available data we cannot rule out hot dust
as a contributing factor.
However, the above discussion shows that the nebular continuum can also be quite
strong at these wavelengths, thus it is hard to separate these
two components in the 3 $-$ 5 $\mu$m wavelength range.

\section{Morphology}

Of the 71 galaxies in our dwarf
sample, 18 are classified in the NASA Extragalactic
Database 
(NED\footnote{The NASA/IPAC Extragalactic Database (NED) is operated by the Jet Propulsion Laboratory, California Institute of Technology, under contract with the National Aeronautics and Space Administration.}) 
as blue compact dwarfs or `compact',
19 as irregular (Im/I0/Irr/IBm/IB/IAB), 3 as Sm, and 3 as Sd, with the rest listed
as other types, such as HII galaxies or starbursts.
In some cases, the classifications in NED differ from those in the SIMBAD 
database\footnote{http:\/\/simbad.u-strasbg.fr\/simbad\/,
operated at CDS, Strasbourg, France}, 
as dwarf classifications are often subjective,
and these catalogs are heterogeneous.

In the left panel of 
Figure 13, we plot the observed [3.6] $-$ [4.5] colors 
of the dwarfs vs.\ log(O/H) + 12
as in Figure 1, however, in Figure 13 the points are color-coded according to the morphological
type in NED.   The blue filled circles are classified as blue compact galaxies or as compact,
while the magenta open diamonds are listed as irregular.   The green crosses are 
galaxies with other classifications.
In
the right panel of Figure 13, we plot [3.6] $-$ [24] vs.\ [3.6] $-$ [4.5]
for the dwarfs as in Figure 2, however, the points are color-coded according
to morphological type rather than metallicity.

Figure 13 shows that 
the galaxies with the reddest [3.6] $-$ [4.5] and [3.6] $-$ [24] colors
are more likely to be classified as blue compact dwarfs
than as irregular.
However, there is considerable scatter in this plot, with some compact galaxies
having blue Spitzer colors.
The observation that the reddest galaxies tend to be classified as compact
agrees with the hypothesis that more compact, and therefore
more UV-intense and/or 
more obscured, star formation
is occurring in these systems.

The reddest systems also show 
evidence for interactions or mergers:
I Zw 18 \citep{vanZee98},
SBSG 0335-052 \citep{pustilnik01},
Haro 11 \citep{ostlin01},
HS 0822+3542 \citep{pustilnik03, corbin06},
NGC 4194
\citep{armus90},
and 
II Zw 40 \citep{bald82}.
This is consistent with the idea that an interaction or merger
has driven gas into the center of the system, causing very
centrally-concentrated star formation.
This may lead to very intense UV fields and/or high optical
depths, causing hotter dust grains, more 24 $\mu$m emission, more reddening
due to extinction, more gaseous continuum emission,
and/or more Br$\alpha$.

In our earlier Spitzer study of more massive interacting galaxies
\citep{smith07}
we did not find a statistically significant
difference between their [3.6] - [4.5] colors and the colors of
normal spirals.   
However, that sample was selected to be early-stage
encounters, with the disks widely separated, 
while some of the dwarfs in the current
sample may be more advanced interactions.

\section{Spitzer Luminosities and Stellar Masses }

For irregular and dwarf elliptical galaxies, a correlation between
blue luminosity and oxygen abundance has been observed
(e.g., \citealp{skillman89}), and interpreted as a mass-abundance relation.
For irregulars, this relation has been extended to the 4.5 $\mu$m luminosity
by \citet{lee06}, who found less scatter at 4.5 $\mu$m than at B, and concluded
that the 4.5 $\mu$m band is a better tracer of stellar mass than the B band.

In the left and right panels of Figure 14, 
we compare the 4.5 $\mu$m
luminosity with log(O/H) + 12
for our sample dwarfs and
normal spirals, respectively.
For consistency with previous work, we use the `monochromatic' luminosity
$\nu$L$_{\nu}$, with frequency 6.67 $\times$ 10$^{13}$ Hz.
The luminosities were calculated using distances from NED calculated with H$_{o}$
= 73 km~s$^{-1}$~Mpc$^{-1}$, and Virgo/Great Attractor/Shapley cluster infall,
except in the case of NGC 6822, where this method fails.  For NGC 6822, we use
the Cepheid distance from \citet{paturel02}.
As expected, the 4.5 $\mu$m luminosities of the dwarf sample are typically
much lower in than for the spirals, but there is some overlap in range.
The points in Figure 14a are color-coded according to their
morphological type in NED, as in Figure 13.

In Figure 14, we include the best-fit line from \citet{lee06}
for their irregular galaxies.
Figure 14 shows that most of our sample galaxies, and all
of our irregulars, lie near the \citet{lee06}
line.  However, a handful of galaxies show large excesses
in 4.5 $\mu$m luminosity for their metallicities.
These include some of the systems noted earlier for having very red
[3.6] $-$ [4.5] colors: SBSG 0335-052, SHOC 391, Harol 11, and I Zw 18.
In contrast, some of the galaxies with red [3.6] $-$ [4.5] colors,
such as II Zw 40, lie close to the \citet{lee06} line.
In addition, a few other galaxies that are less extreme in [3.6] $-$ [4.5]
also show up as discrepant in Figure 14: SHOC 567, Tol 2138-405, and UM 420.
Tol 2138-405 and SHOC 567, like the redder galaxies, have peculiar morphologies
indicative of interactions \citep{telles97, kniazev04}.   UM 420 lies
behind a foreground elliptical galaxy \citep{salzer89}, thus the published Spitzer
flux may be confused.

Figure 14 shows that for extreme starbursts, compact systems, and/or 
interacting systems, the 4.5 $\mu$m luminosity may not be a good tracer
of stellar mass.  This may be in part due to non-stellar contributions
in the 4.5 $\mu$m band.   Alternatively, an interaction or merger
may have driven unprocessed gas into the central region of these galaxies,
lowering their metallicity relative to their
stellar mass
(e.g., \citealp{rupke08}).

In Figure 15, we plotted the 3.6 $\mu$m luminosity vs.\ oxygen abundance
for both the dwarfs and the spirals.   Again, a correlation is seen,
in that higher luminosity systems tend to have higher metallicities.
However, as in Figure 14,
there are some galaxies with too high luminosities
for their metallicities.   These are the same galaxies that are
discrepant in Figure 14, thus these galaxies
may also have large non-stellar excesses at 3.6 $\mu$m,
or may have lowered metallicities due to interactions or mergers.

\section{Discussion}

For the dwarf galaxies in our sample, we find that lower metallicity
systems tend to have redder [3.6] $-$ [4.5] colors than higher metallicity
systems, in that systems with log(O/H) + 12 $\le$ 8.2
tend to be redder on average.
Comparison to stellar population synthesis models shows that 
the observed colors of many dwarfs are too red to
be accounted for solely by unobscured starlight.

For all of the galaxies in our sample 
with sufficient data available,
a combination of 
Br$\alpha$ emission, reddening by dust, and nebular emission 
can account for 
the observed 4.5 $\mu$m excess.   
Nebular continuum
emission, which is often neglected in discussions of the Spitzer colors
of galaxies, may contribute to both the 3.6 $\mu$m and 4.5 $\mu$m
bands, and in extreme cases can dominate the observed fluxes in these bands.
As seen in Figure 2, 4.5 $\mu$m excess is correlated with the mass-normalized 24 $\mu$m
luminosity and therefore with more intense UV fields.  In lower metallicity systems,
the 4.5 $\mu$m excess is larger for the same [3.6] $-$ [24] color,
consistent with both stronger Br$\alpha$ emission
and more nebular continuum for lower abundances.

Another factor that is sometimes ignored
is reddening by dust extinction in the Spitzer bands, which
can be quite large in some situations.
Some of our sample galaxies show deep silicate absorption features
and/or large Br$\alpha$/Br$\gamma$ ratios, in spite of the small
extinctions implied by H$\alpha$/H$\beta$ or optical colors.  
In these
cases, reddening of the starlight due to dust extinction may
be quite important.
The situation in these very obscured dwarfs may resemble that in 
the starburst elliptical galaxy NGC 1377, which has a deep silicate
absorption feature and a high implied optical depth \citep{roussel06}.
NGC 1377 may contain
a very young highly obscured compact starburst
\citep{roussel06}.
Like our extreme dwarfs, NGC 1377 
has a very red [3.6] $-$ [4.5] color (see Figure 16 in
\citealp{smith07}).

Given the uncertainties in the data and models, we cannot rule out
an additional contribution to the 4.5 $\mu$m flux from 
dust 
heated to high temperatures by very intense UV fields.
This dust continuum may be the short-wavelength
extension of the VSG continuum seen at longer wavelengths
in ISO and Spitzer spectra of Galactic HII regions and 
star forming regions in 
dwarf galaxies.
At the present time, the nature of the grains responsible for 
this 4.5 $\mu$m dust emission is unclear.   
According to the dust models of 
\citet{dl07},
the grains responsible for the continuum
at 6 $-$ 30 $\mu$m have 
sizes $\sim$ 15$-$40$\AA$, and may be
large PAHs (N$_C$ $\sim$ 2000 $-$ 3 $\times$ 10$^4$).
\citet{dl07}
concluded that in most cases
this continuum emission is due to single-photon heating, however,
in the case of very intense UV radiation fields,
a particle may not
cool completely before absorbing another photon.
In this case,
higher temperatures are reached, 
and the peak of the VSG spectrum moves to shorter
wavelengths.
In sources with strong
4.5 $\mu$m continuum, the particles may be larger and/or hotter,
and possibly may be in thermal equilibrium with a hard intense
UV field.

It is unclear how the grains that produce the strong 6 $-$ 30 $\mu$m
continuum in HII regions are related to the carriers of the 
$\sim$2 $-$ 3.7 $\mu$m
continuum 
found by 
\citet{sellgren83}
in reflection nebulae, which was
attributed to small 
grains ($\sim$10\AA) stochastically
heated 
to $\sim$1000K 
by UV or 
optical photons 
\citep{sellgren84}.
The ambient UV fields in these two environments are quite different.
\citet{an03}
suggest that the near-infrared continuum-producing particles 
in reflection nebulae may be PAHs that
are larger, more ionized,
and/or more dehydrogenated than those that produce the 3.3 $\mu$m PAH
feature, or
may require higher energies for excitation.

Interestingly, a 4.5 $\mu$m excess has also 
been found for the diffuse Galactic interstellar medium.
\citet{dl07}
found that the 
\citet{flagey06}
Spitzer broadband measurements of diffuse Milky Way dust
are consistent with standard dust models, 
except for the 4.5 $\mu$m flux, which is too high by a factor of
$\sim$ 1.6.
\citet{flagey06}
and 
\citet{dl07}
conclude that there is a
`near-infrared continuum' which contributes at 4.5 $\mu$m and 3.6 $\mu$m.
Furthermore, 
\citet{li01}
compared their dust models
to high Galactic latitude emission as seen by DIRBE, and 
found a 4.9 $\mu$m excess 
($\sim$2.5$\times$).
They suggested 
an additional form of opacity at these wavelengths, for example, carbon chains
(e.g., Allamandola et al.\ 1999).
It is unclear whether the particles that produce this 4 $-$ 5 $\mu$m excess
in the diffuse ISM are the same as those that are responsible for
the excess in dwarf galaxies.

As noted by \citet{engel05} and \citet{draine07},
there appears to be a sharp change in the 8/24 $\mu$m flux ratio
of
galaxies at log(O/H) + 12 = 8.2 ($\sim$1/3 solar), 
in that this ratio abruptly
decreases at this metallicity, perhaps because of weakening
of the 7.7 $\mu$m PAH feature at low metallicities.
We see an abrupt change in the [3.6] $-$ [4.5] color near this same
metallicity, suggesting that the 4.5 $\mu$m excess and the PAH
deficiency are related.
One possibility is that at this metallicity threshold the UV field hardens
sufficiently to more efficiently destroy PAHs, and at the same time 
produces more Br$\alpha$, more infrared nebular continuum, and/or
more effectively excite the carriers of 4.5 $\mu$m dust emission.
A second possibility is that the destruction of the carriers of
the PAH emission features produces
the particles that are responsible for a 4.5 $\mu$m dust continuum.

A third possibility is that 
there is an abrupt
change in the chemistry of the interstellar matter
at this metallicity, leading to both
a deficiency in PAHs and to a larger percentage
of grains that are more effective at radiating
in the 4.5 $\mu$m regime. 
For example, the galaxian chemical evolution model
of \citet{galliano08} shows a time delay
between the production of silicate dust grains in supernovae
and the formation of carbon-rich
material
in asympotic giant branch stars, which
can 
account for the relative deficiency of PAHs at low metallicities.
As noted earlier, the galaxies in our sample
with extremely red [3.6] $-$ [4.5]
colors have infrared spectral energy distributions that peak
near 24 $\mu$m, while other low metallicity systems do not.
\citet{engelbracht08}
suggest that 
the particles responsible for far-infrared/submillimeter
emission are deficient in these galaxies, or 
are inefficiently heated.  Thus the composition of the interstellar
dust in these galaxies may be different from other low metallicity
systems.

For all of our galaxies with sufficient
optical data except one, population synthesis modeling
cannot rule out that most of the 3.6 $\mu$m Spitzer flux is due to
the stars seen at optical wavelengths.  The exception,
SBSG 0335-052, either has a highly obscured stellar population
unseen at optical and near-infrared wavelengths, or a strong nebular
or hot dust continuum that contributes significantly
at 3.6 $\mu$m as well as at 4.5 $\mu$m.
A similar situation applies for 
a star forming clump in 
the tidal tail of the interacting galaxy pair Arp 285.  For this
clump, population synthesis based on optical broadband data indicates an
excess at both 
3.6 $\mu$m and 4.5 $\mu$m flux above the inferred stellar continuum
\citep{smith08}.
As with SBSG 0335-052, this may be due to either a strong nebular continuum,
hot dust being present
at 3.6 $\mu$m, or high obscuration in the optical.

\section{Conclusions}

For a large sample of dwarf galaxies, we show that the Spitzer
broadband [3.6] $-$ [4.5]
colors depend upon metallicity, in that lower metallicity systems
(log(O/H) $-$ 12 $<$ 8.2) show redder colors on
average than higher metallicity systems, but with much scatter. 
Comparison with stellar population synthesis models shows that
starlight alone cannot account for the range of observed colors,
even when accounting for variations in age and metallicity.
Many galaxies have an excess at 4.5 $\mu$m compared to
that expected from stars.
In some galaxies, Br$\alpha$ contributes significantly to the 
observed 4.5 $\mu$m flux.  In other systems, the nebular continuum
plays an important role in reddening the [3.6] $-$ [4.5] color, and 
in extreme cases may dominate the observed flux at both 3.6 $\mu$m and 4.5 $\mu$m. 
The lower metallicities in dwarfs hardens the UV field, increasing
the Br$\alpha$ luminosity and the gaseous continuum in the infrared.
In addition, high obscuration can also redden the [3.6] $-$ [4.5] color in some cases.
The combined effects of these three factors can account for the observed [3.6] $-$ [4.5]
colors of all the galaxies in our sample that have sufficient data available.
Given the uncertainties, we cannot rule out an additional component 
from hot dust at 4.5 $\mu$m.   
Harder UV fields may
increase the average dust temperature, increasing dust contributions
at 4.5 $\mu$m.
However, for our sample galaxies it does not appear to be
the dominant factor.

The galaxies with the largest 4.5 $\mu$m
excess are classified as blue compact dwarfs, and show
evidence of recent interactions and mergers with other galaxies.
Such encounters may produce very concentrated star 
formation, causing very intense UV fields and/or high extinctions.
For these galaxies, the 4.5 $\mu$m luminosity may not be a good
tracer of stellar mass.

A similar analysis was done for a sample of nearby normal spirals.
This showed that contributions from Br$\alpha$
and the nebular continuum to the 4.5 $\mu$m fluxes of normal spirals
are generally small, as is reddening of the [3.6] $-$ [4.5] starlight
by dust.
This is consistent with their observed [3.6] $-$ [4.5] colors of $\sim$ 0.0.
For normal spirals, these bands appear to be dominated by starlight.

\acknowledgements

We thank the Spitzer team for making this research possible.
This research was supported by 
NASA LTSA grant NAG5-13079.
We thank Mark Giroux and Curt Struck for helpful communications.
We also thank the anonymous referee for many helpful suggestions that
greatly improved the paper.
This research has made use of the NASA/IPAC Extragalactic Database 
(NED) which is operated by the Jet Propulsion Laboratory, California 
Institute of Technology, under contract with the National Aeronautics and Space Administration.
We acknowledge the use of the HyperLeda database
at 
http://leda.univ-lyonl.fr




\clearpage




\begin{figure*}
\includegraphics[width=3.1in]{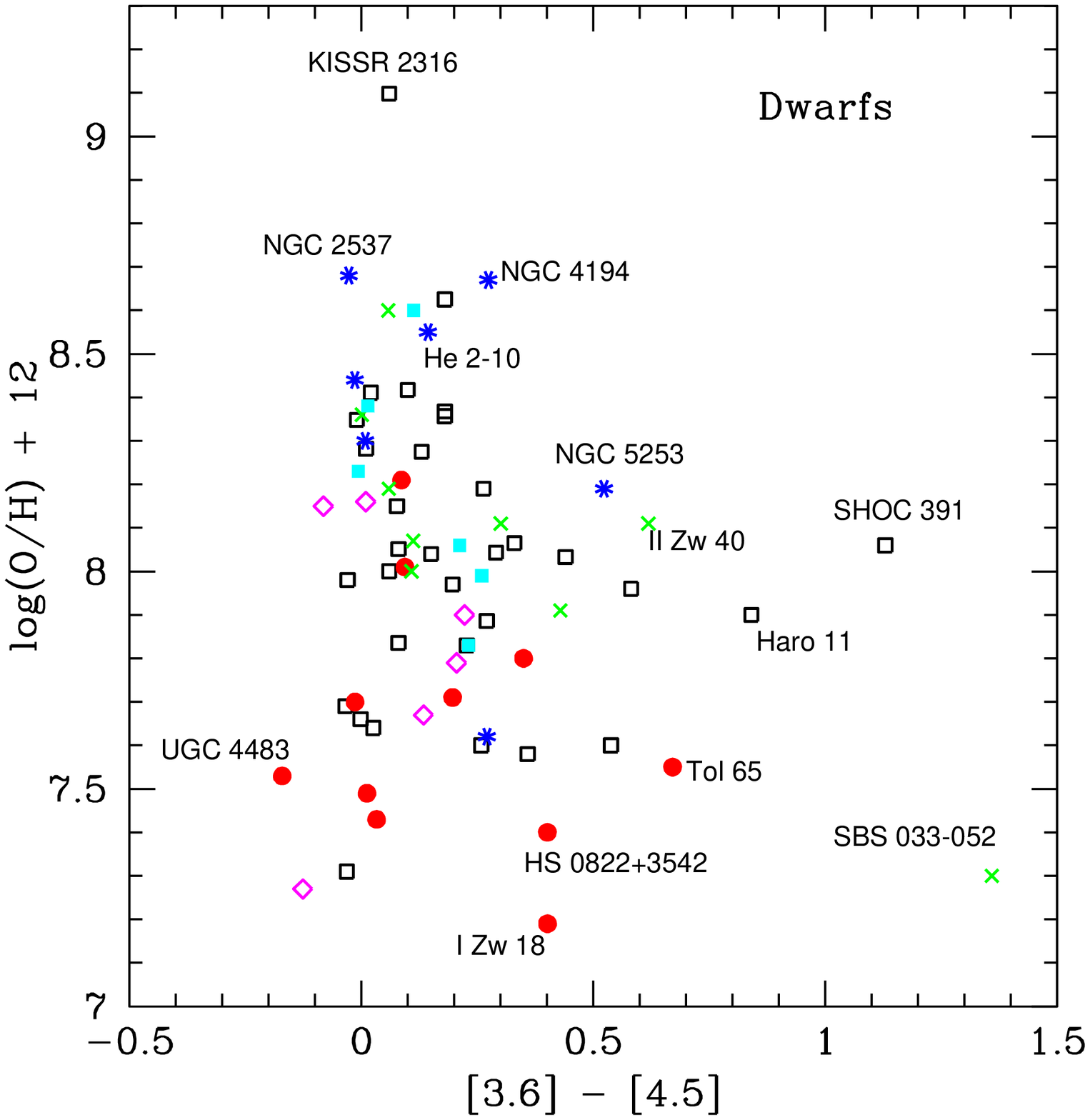}
\includegraphics[width=3.1in]{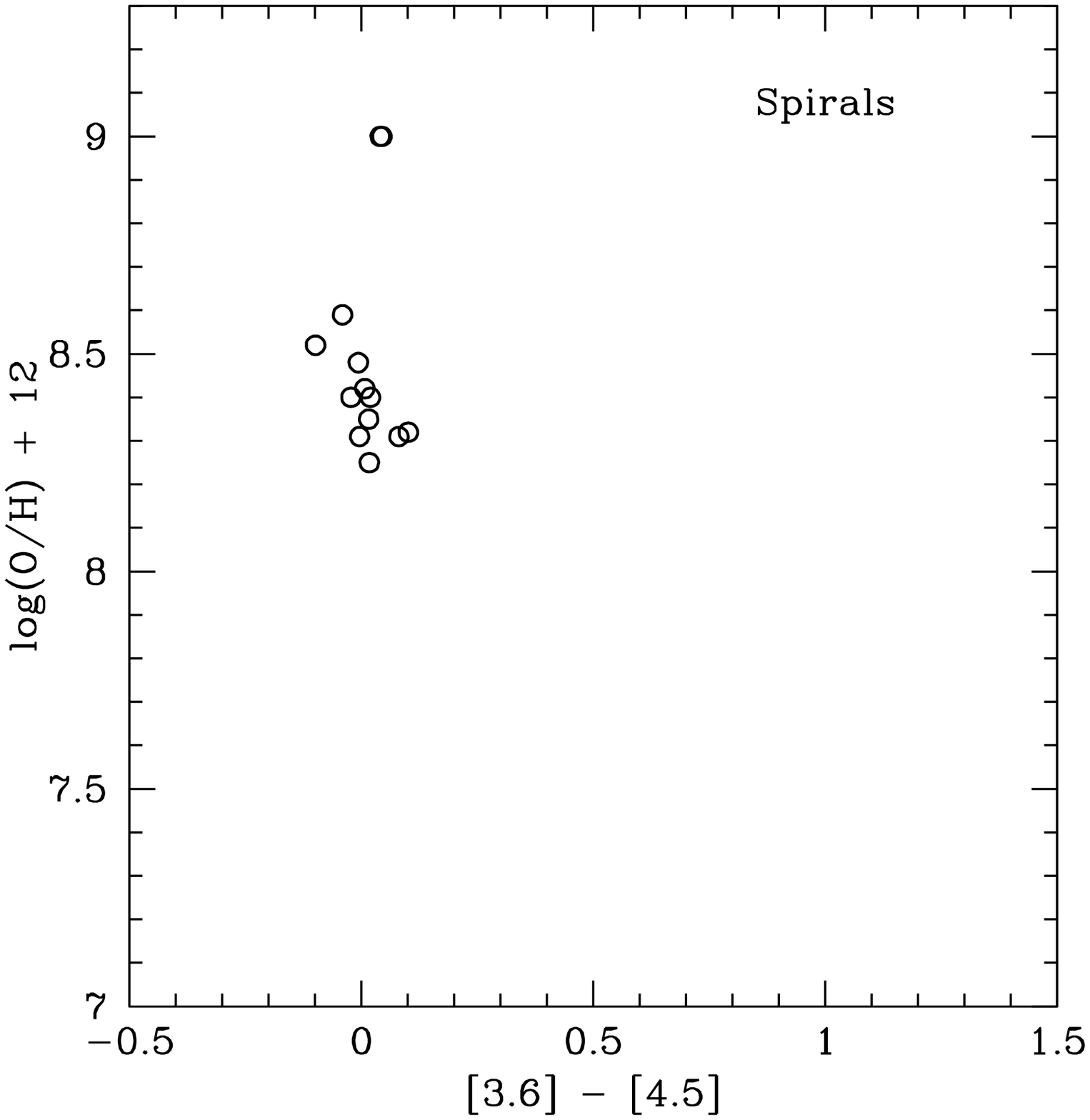}
\caption{Left: A plot of log(O/H) + 12 vs.\ [3.6] $-$ [4.5] for the sample
dwarf galaxies.  The different colors represent different HI mass to stellar
mass ratios, as indicated by the HI magnitude 
m(HI) minus the 3.6 $\mu$m magnitude [3.6].   
The HI magnitude is defined
as in Paturel et al.\ (2003).
The magenta 
open diamonds are the most HI-rich
systems, with 
mag(HI) - [3.6] $<$ 1.3.  Red filled circles have
1.3 $<$ mag(HI) - [3.6] $\le$ 3.0, green crosses
3.0 $<$ mag(HI) - [3.6] $\le$ 4.1, cyan filled squares
4.1 $<$ mag(HI) - [3.6] $\le$ 5.5, and blue asterisks
5.5 $<$ mag(HI) - [3.6].
Galaxies marked in black open squares do not have HI data available.
Typical uncertainties on the Spitzer [3.6] $-$ [4.5] colors of galaxies
are $\sim$0.05 (Smith et al.\ 2007).
Some of the galaxies are labeled.
Right: A similar plot for the comparison sample of `normal' spiral galaxies.
\label{fig1}}
\end{figure*}

\clearpage


\begin{figure*}
\includegraphics[width=3.1in]{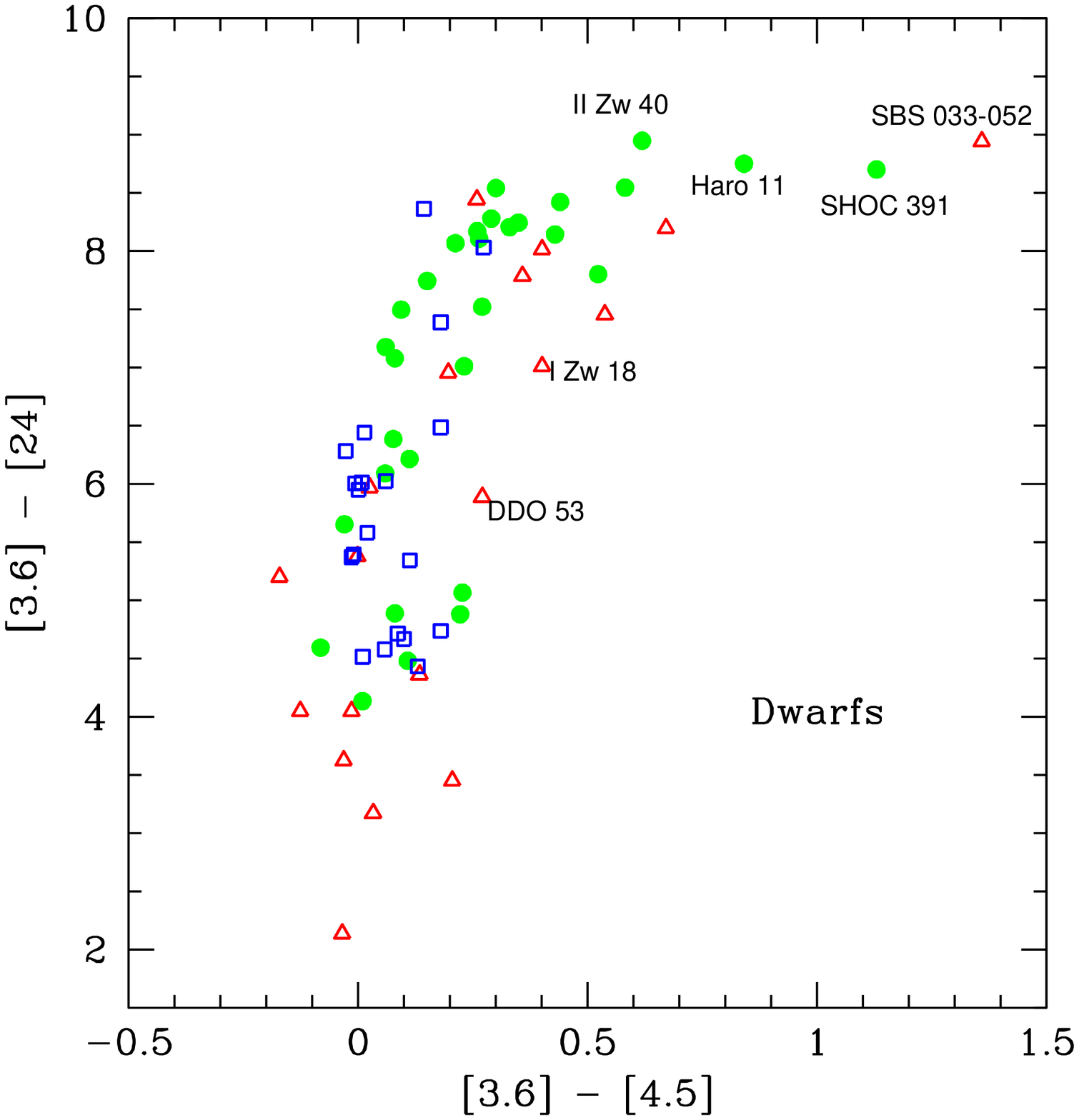}
\includegraphics[width=3.1in]{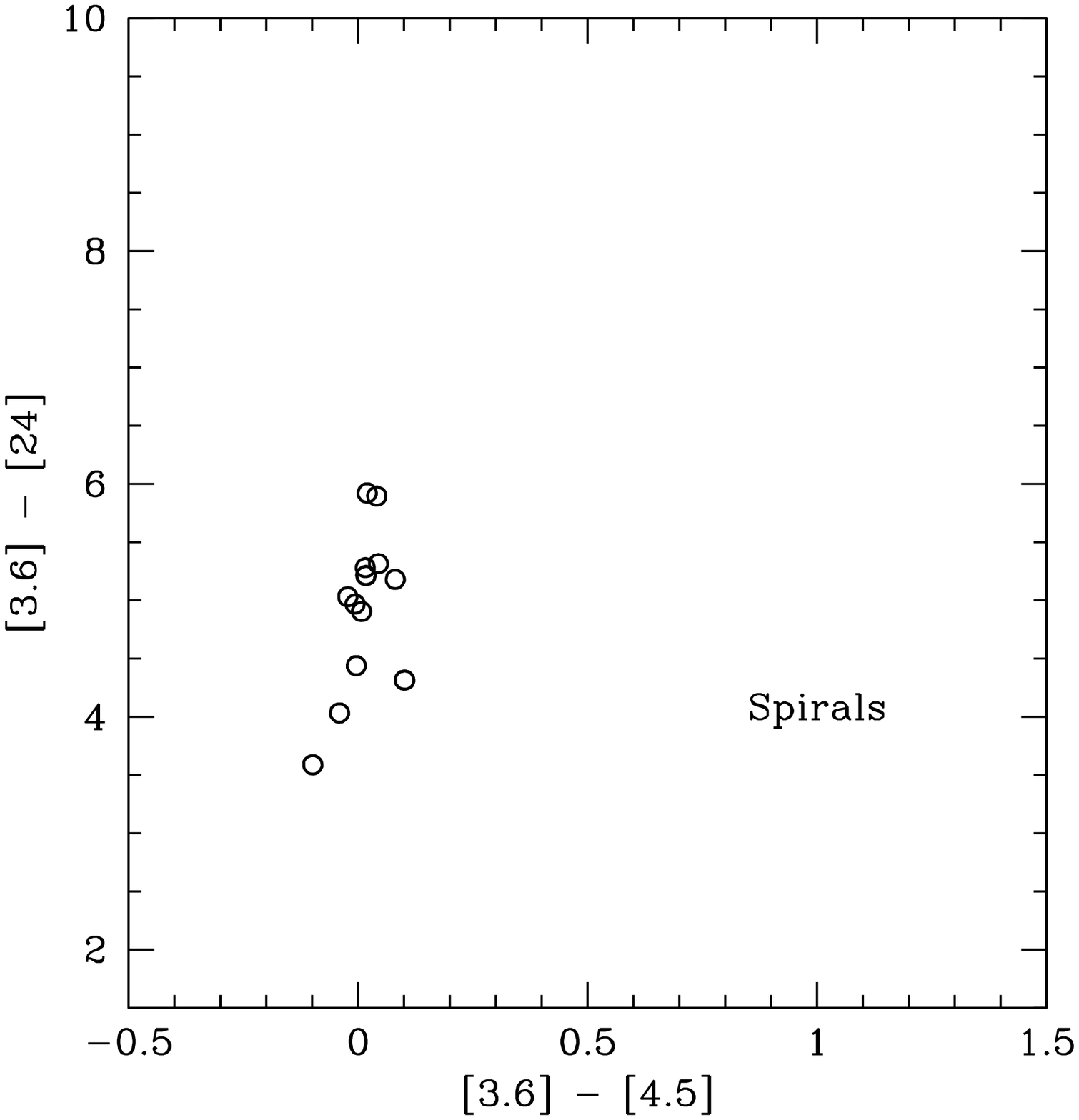}
\caption{Left: The [3.6] $-$ [24] vs.\ [3.6] $-$ [4.5] colors for the sample
dwarf
galaxies.  The different colors represent different metallicity ranges.
The red open triangles are log(O/H) + 12 $<$ 7.8, the green filled circles
7.8 $\le$ log(O/H) + 12 $\le$ 8.2, and the blue open squares
log(O/H) + 12 $>$ 8.2.
Right: A similar plot for the `normal spirals' sample.
\label{fig2}}
\end{figure*}


\begin{figure*}
\includegraphics[width=3.1in]{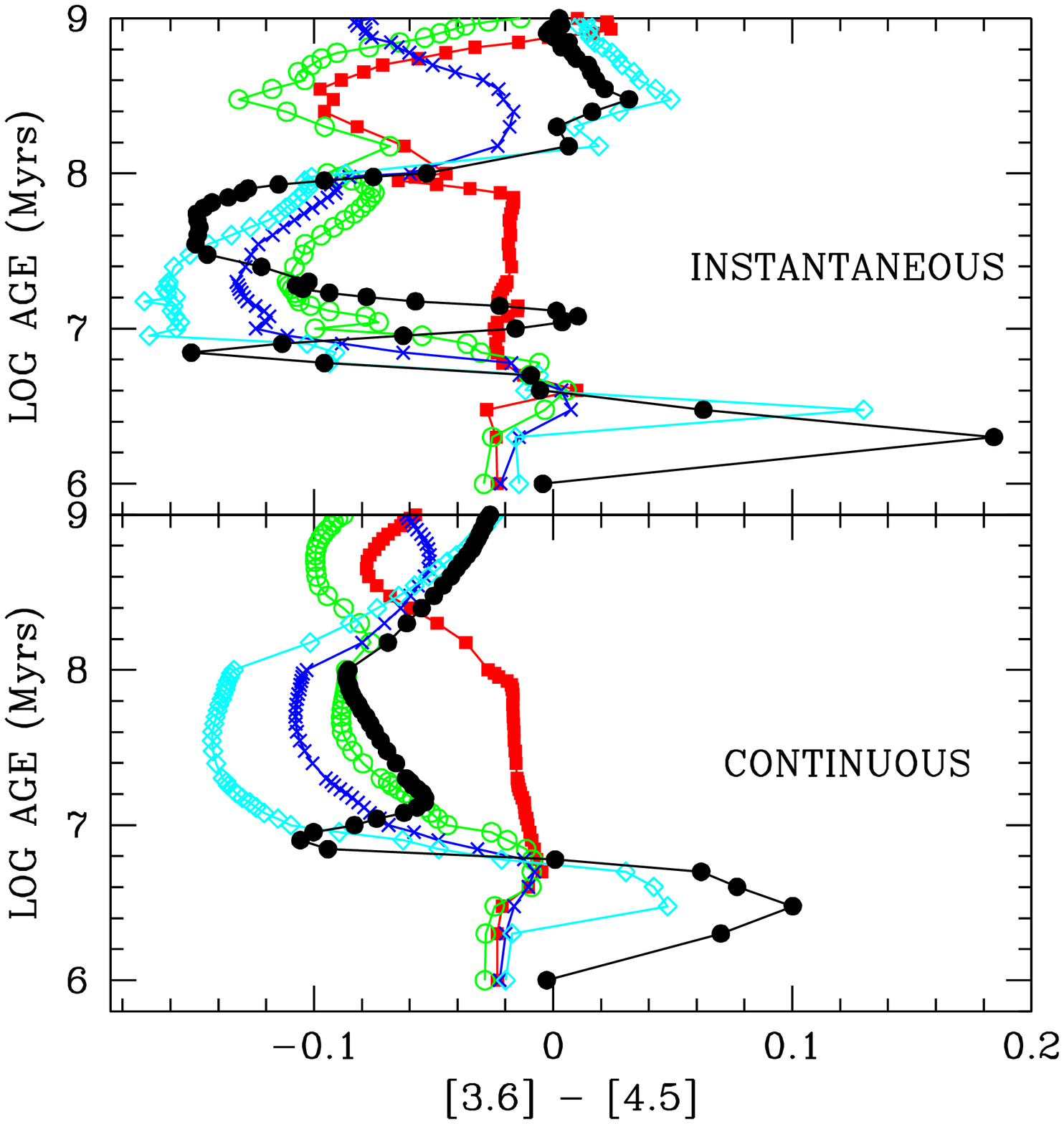}
\includegraphics[width=3.1in]{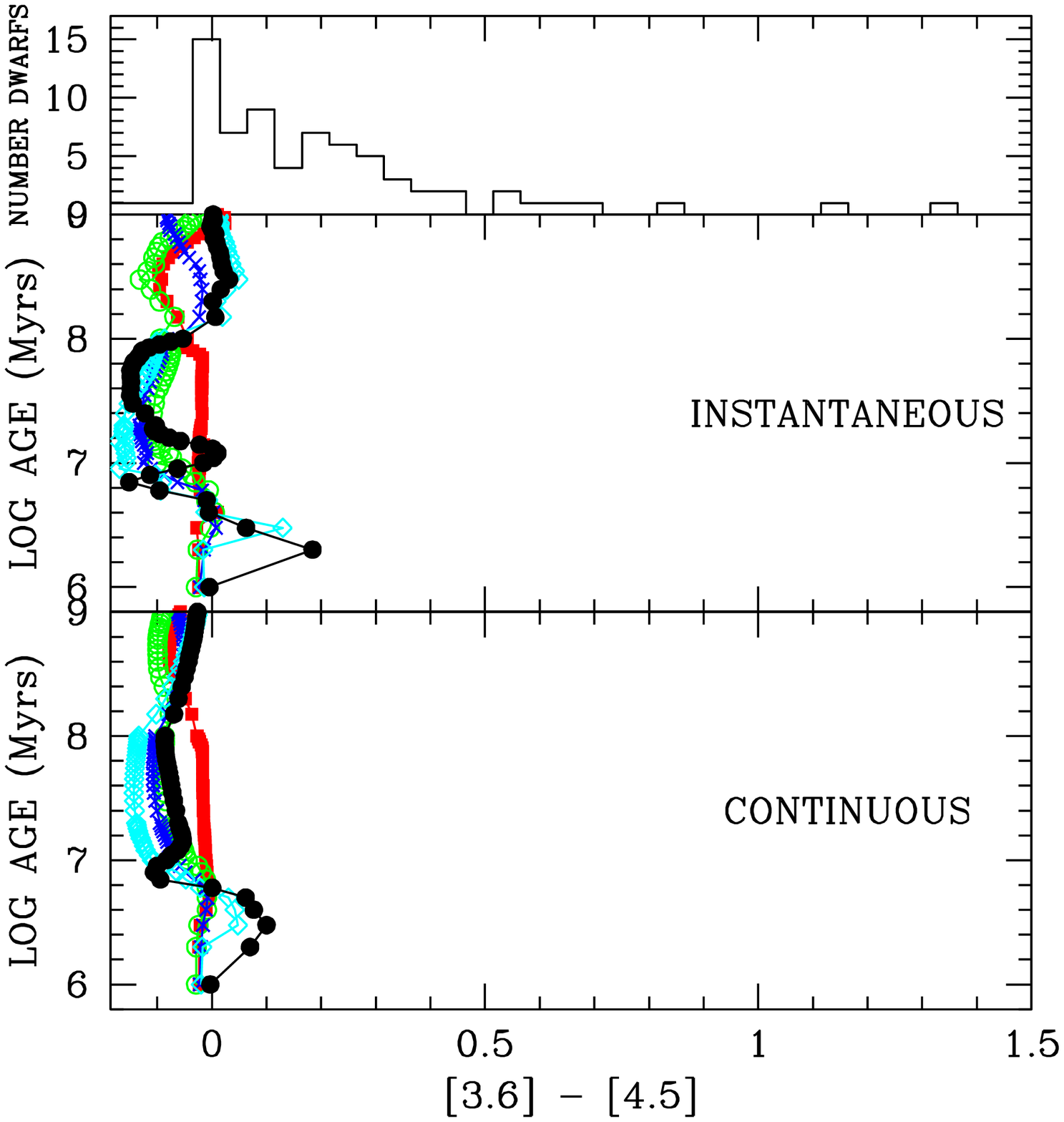}
\caption{Left: The [3.6] $-$ [4.5] colors for the Starburst99 models
as a function of age.
Only starlight is included in these models; no contributions from
the interstellar medium are included.
The top panel shows the results for the
instantaneous burst models; the lower panel gives the 
continuous star formation models.   
The different colors
and symbols
represent different metallicities, with red (filled squares)
being
1/50 Z$_{\odot}$, green (open circles)
1/5 Z$_{\odot}$, blue (crosses) 1/2.5 Z$_{\odot}$,
cyan (open diamonds) 1 Z$_{\odot}$, and black (filled circles)
2.5 Z$_{\odot}$.
Right panel: lower two plots are
the same as the left panel, except with an expanded x-axis,
for comparison with Figures 1 and 2.
The top panel on the right shows a histogram of the [3.6] $-$ [4.5]
colors for the dwarf sample, for comparison with the models.
\label{fig3}}
\end{figure*}

\begin{figure*}
\plotone{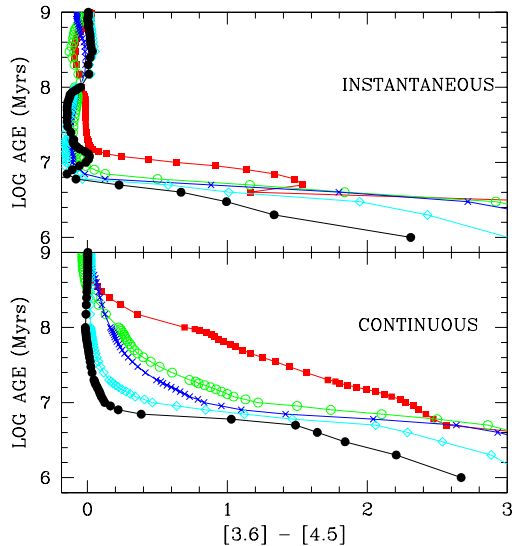}
\caption{Model [3.6] $-$ [4.5] colors for stellar population models
as a function of age, as in Figure 3, except with Br$\alpha$ added to the
4.5 $\mu$m band.
The symbols and colors are as in Figure 3.
Note that the scale on the x-axis extends further than in Figure 3b.
\label{fig4}}
\end{figure*}

\begin{figure*}
\includegraphics[width=3.1in]{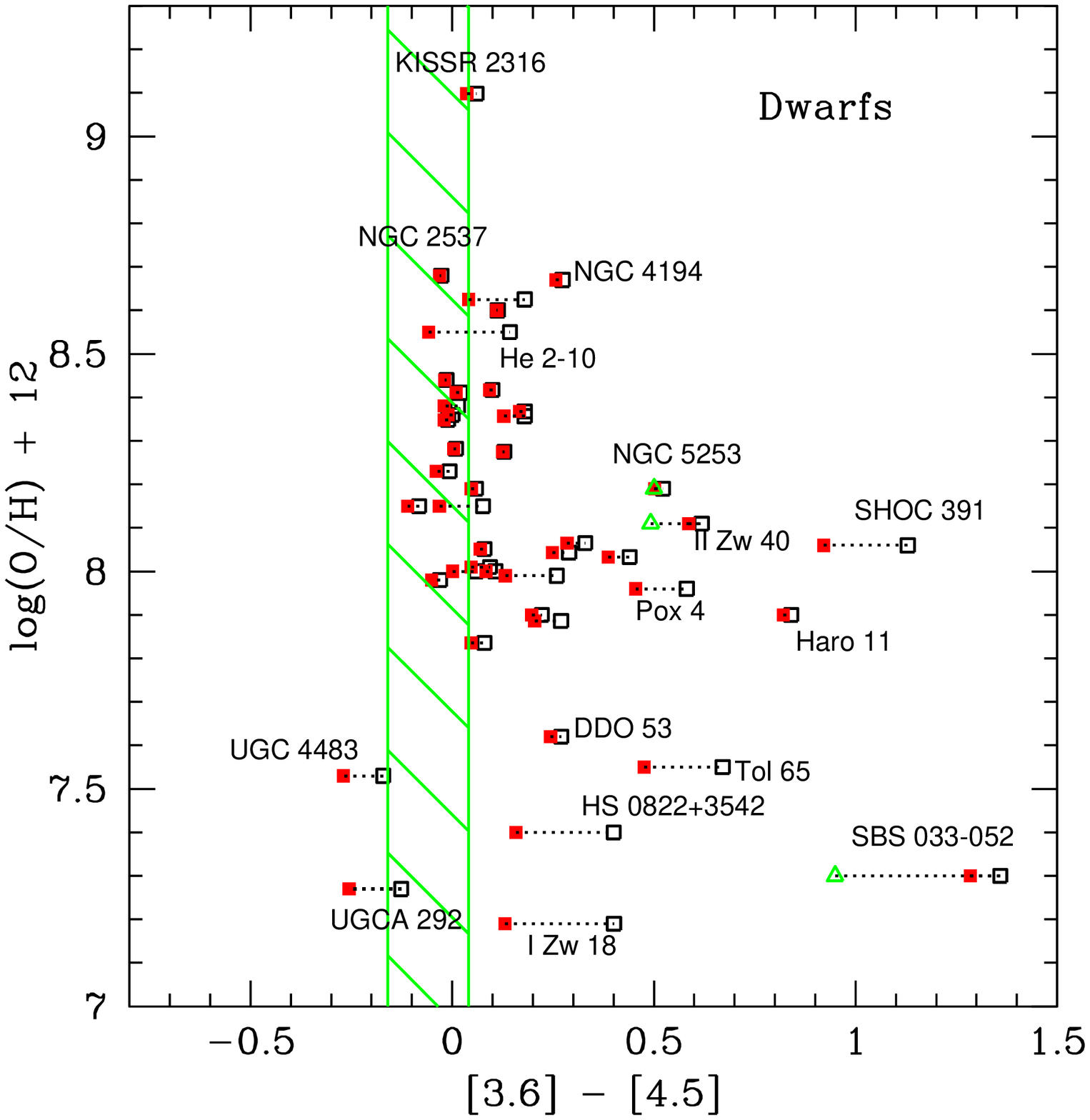}
\includegraphics[width=3.1in]{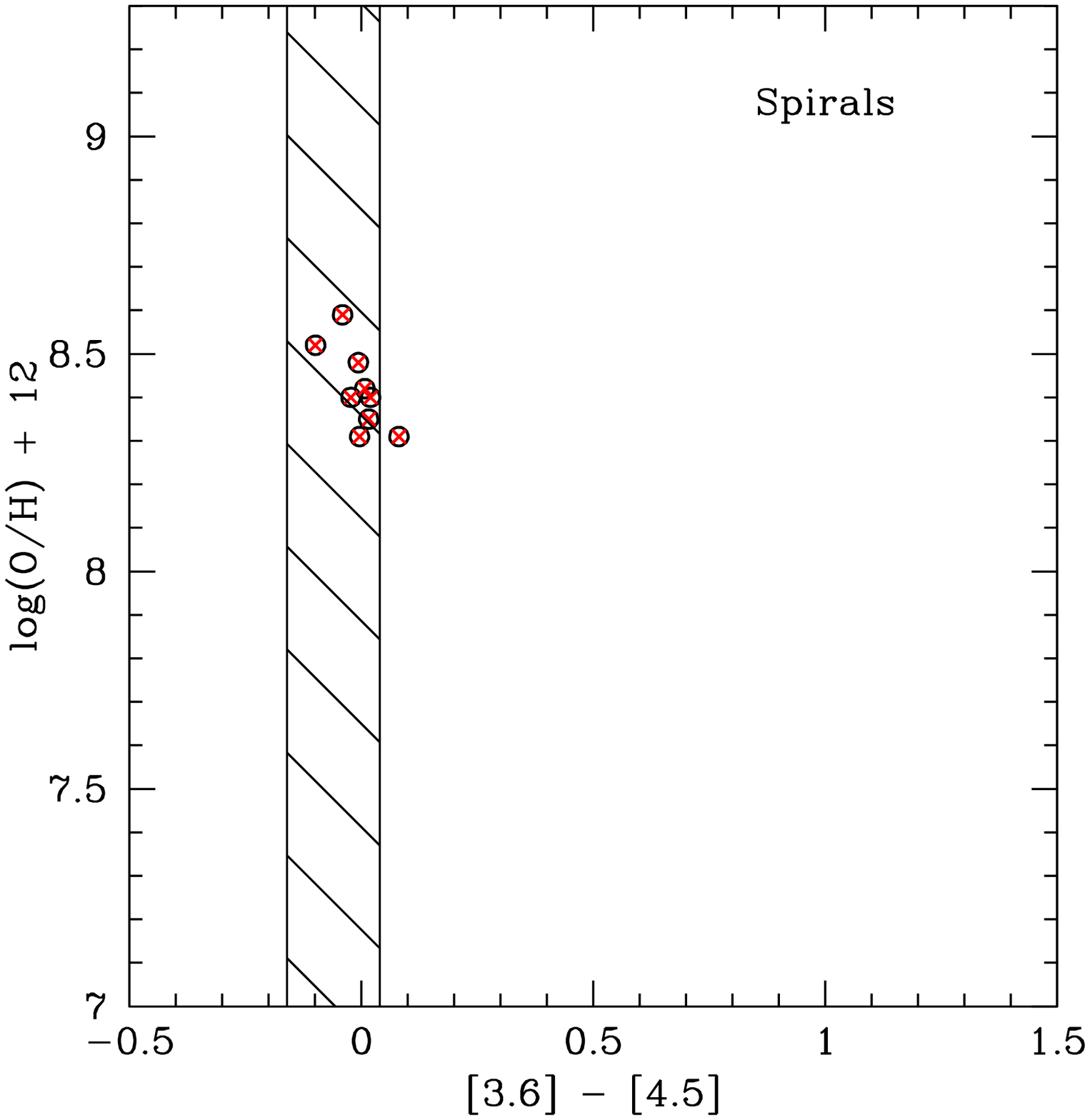}
\caption{Left: The oxygen abundance of the subset of our sample dwarf galaxies
with published H$\alpha$ fluxes, plotted against the observed
[3.6] $-$ [4.5] color (open black squares) and the [3.6] $-$ [4.5]
color after correction for Br$\alpha$.
The filled red squares are the Br$\alpha$-corrected colors after
correction using H$\alpha$; the open green triangles
are after correction using measured Br$\alpha$ or Br$\gamma$ fluxes.
The Br$\alpha$-corrected datapoints are connected to the observed datapoints
by dotted lines.
The expected range for unextincted starlight is shown by hatch marks.
Right: A similar plot for the sample spirals, where the open circles are the uncorrected
points, and the red crosses are after correction for Br$\alpha$ using total H$\alpha$ fluxes
from \citet{kennicutt03}
and Pa$\alpha$/H$\alpha$ ratios for the inner 50$''$ from \citet{calzetti07}.
Only galaxies with total H$\alpha$ fluxes and published
Pa$\alpha$/H$\alpha$ values are plotted.
\label{fig5}}
\end{figure*}

\begin{figure*}
\includegraphics[width=3.1in]{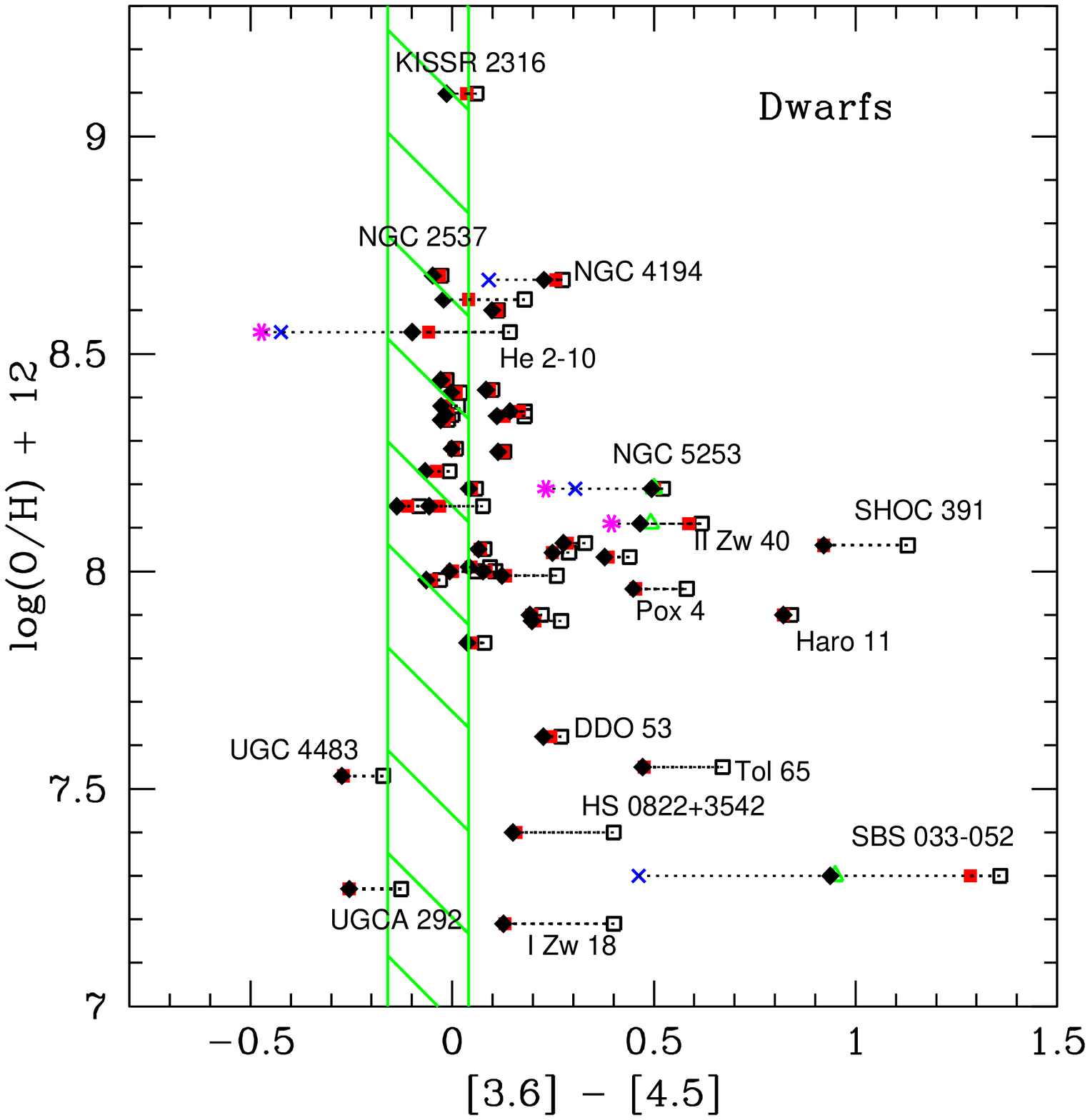}
\includegraphics[width=3.1in]{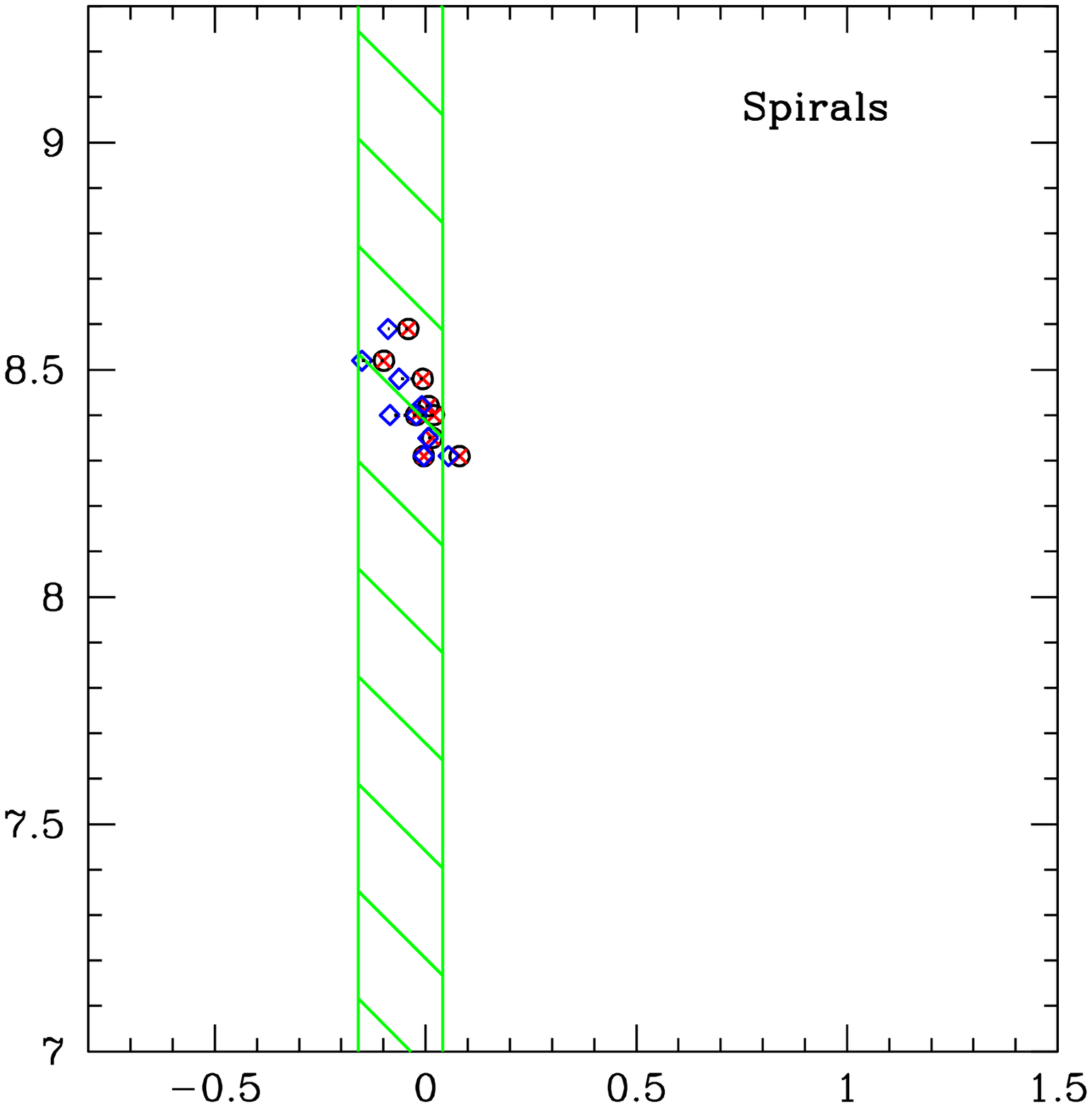}
\caption{Left: The same as Figure 5a, except with points added
to show an additional correction for reddening of the starlight due to 
extinction by dust.   Blue crosses represent points corrected
using extinction estimates from the 9.7 $\mu$m silicate absorption
feature.  Magenta asterisks show points corrected using
Br$\alpha$/Br$\gamma$ ratios, and black filled diamonds show points
corrected for reddening due to dust extinction
using H$\alpha$/H$\beta$ ratios.
The correction for extinction has been added to the correction
for Br$\alpha$ in this plot.
The extinction-corrected points are connected to the Br$\alpha$-corrected
points by dotted lines.
Right: The same as the left panel, except for the spiral sample.
The blue open diamonds show the colors after correction for
reddening of the starlight using the Pa$\alpha$/H$\alpha$ ratio
of \citet{calzetti07}.
\label{fig6}}
\end{figure*}

\begin{figure*}
\plotone{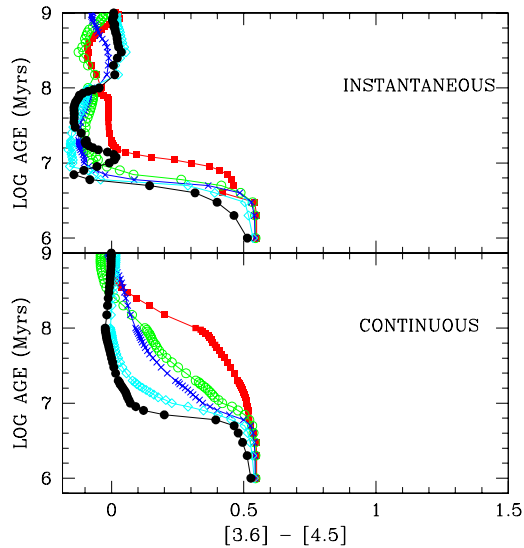}
\caption{Model [3.6] $-$ [4.5] colors as a function of age as in Figure 3,
except with the nebular continuum added. 
No Br$\alpha$ or reddening by dust extinction is included.
The symbols are as in Figure 3.
The x-axis range has been set to match that in Figures 1, 2, and 3b.
\label{fig7}}
\end{figure*}

\begin{figure*}
\plotone{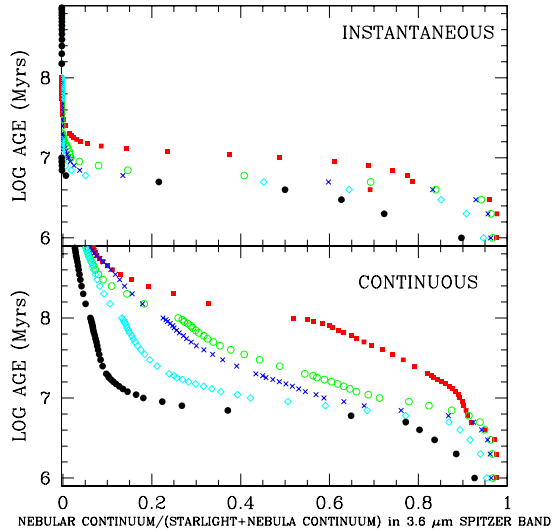}
\caption{The results of Starburst99 models
showing the ratio of the flux in the 3.6 $\mu$m band from the nebular continuum
to that from starlight plus the nebular continuum.  
The different colors
and symbols
represent different metallicities, with red (filled squares)
being
1/50 Z$_{\odot}$, green (open circles)
1/5 Z$_{\odot}$, blue (crosses) 1/2.5 Z$_{\odot}$,
cyan (open diamonds) 1 Z$_{\odot}$, and black (filled circles)
2.5 Z$_{\odot}$.
\label{fig8}}
\end{figure*}

\begin{figure*}
\plotone{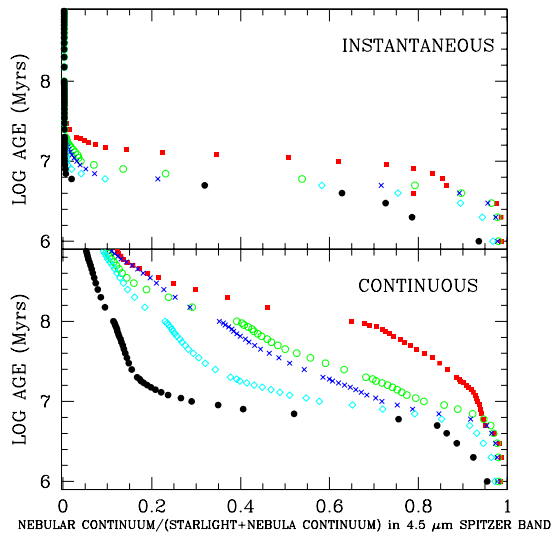}
\caption{The results of Starburst99 models
showing the ratio of the flux in the 4.5 $\mu$m band from the nebular continuum
to that from starlight plus nebular continuum.  
The symbols are as in Figures 3, 4, 7, and 8.
No Br$\alpha$ is included.
\label{fig9}}
\end{figure*}

\begin{figure*}
\includegraphics[width=3.1in]{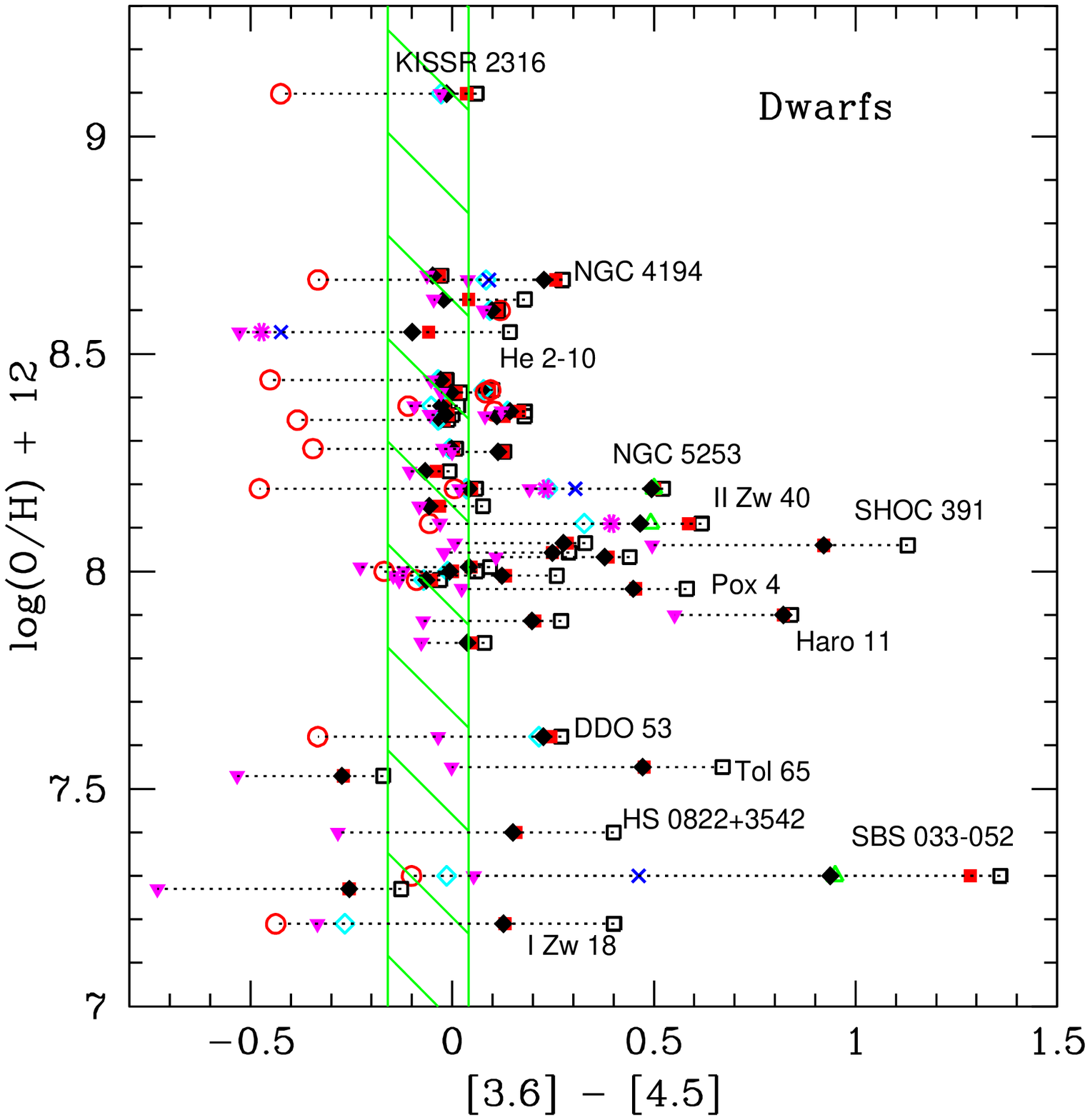}
\includegraphics[width=3.1in]{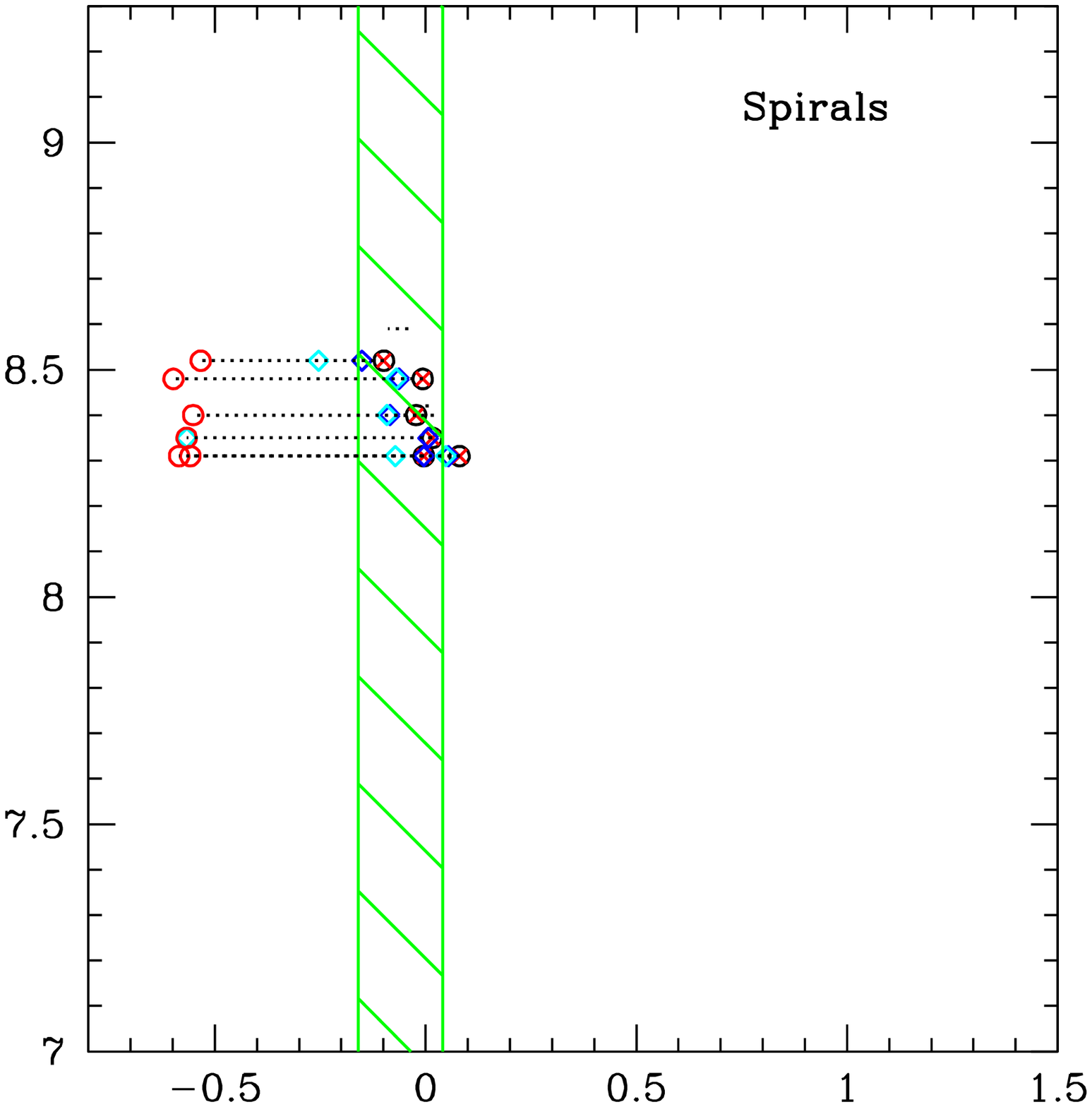}
\caption{As in Figure 6, except with points added to show an additional correction
for reddening due to the nebular continuum.
The left panel shows the results for the dwarfs, the right the spirals.
The magenta filled upside-down triangles show the nebular continuum-corrected
points, calculated using ages determined from the H$\beta$ equivalent width.
The cyan diamonds show the nebular continuum-corrected points calculated
using the best-fit age from the broadband optical data.  The red open circles shows
the nebular-corrected values, calculated using the best fit age from the
broadband data, minus the 1$\sigma$ uncertainty to
the age.
The correction for the nebular continuum has been added to the corrections
for Br$\alpha$ and extinction in this plot.
\label{fig10}}
\end{figure*}

\begin{figure*}
\plotone{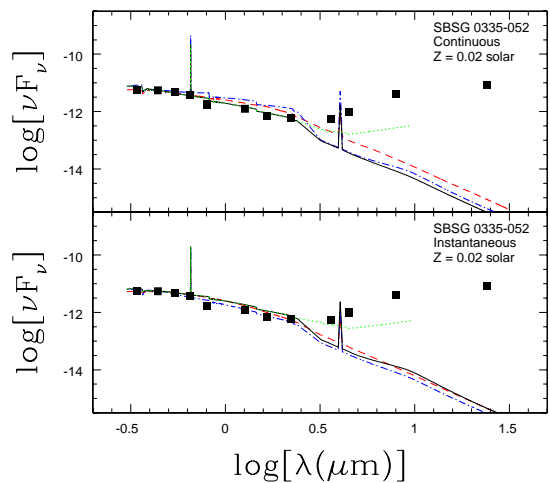}
\caption{The optical-mid-infrared spectral energy distributions
of 
SBSG 0335-052.
The black solid curve is
the best-fit Starburst99 model,
while the blue dot-dashed curve shows the lower limit to the age, and the
red dashed curve the upper limit to the age.
The upper and lower limits curves are plotted with their best-fit extinctions.
The green dotted curve shows the best-fit curve including the nebular continuum
in the Spitzer bands.   The other curves only include starlight in the Spitzer bands;
for these curves, 
the nebular continuum was only included 
in the optical and near-infrared.
The lower panel shows the instantaneous burst models, while the continuous
models are shown in the upper panel.
These models were calculated using a metallicity of 1/50th solar.
The models are scaled to the V band optical flux.
The H$\alpha$ and Br$\alpha$ lines are plotted for comparison,
but no other emission lines.
The optical, near-infrared, and Spitzer data are from
\citet{pustilnik04}, \citet{vanzi00}, and \citet{engelbracht08},
respectively.
The best-fit ages are 37 $\pm$ $^{120}_{28}$ Myrs for the
continuous burst model, and 6 $\pm$ $^{6}_{3}$ Myrs for an
instantaneous burst.
\label{fig11}}
\end{figure*}

\begin{figure*}
\includegraphics[width=3.1in]{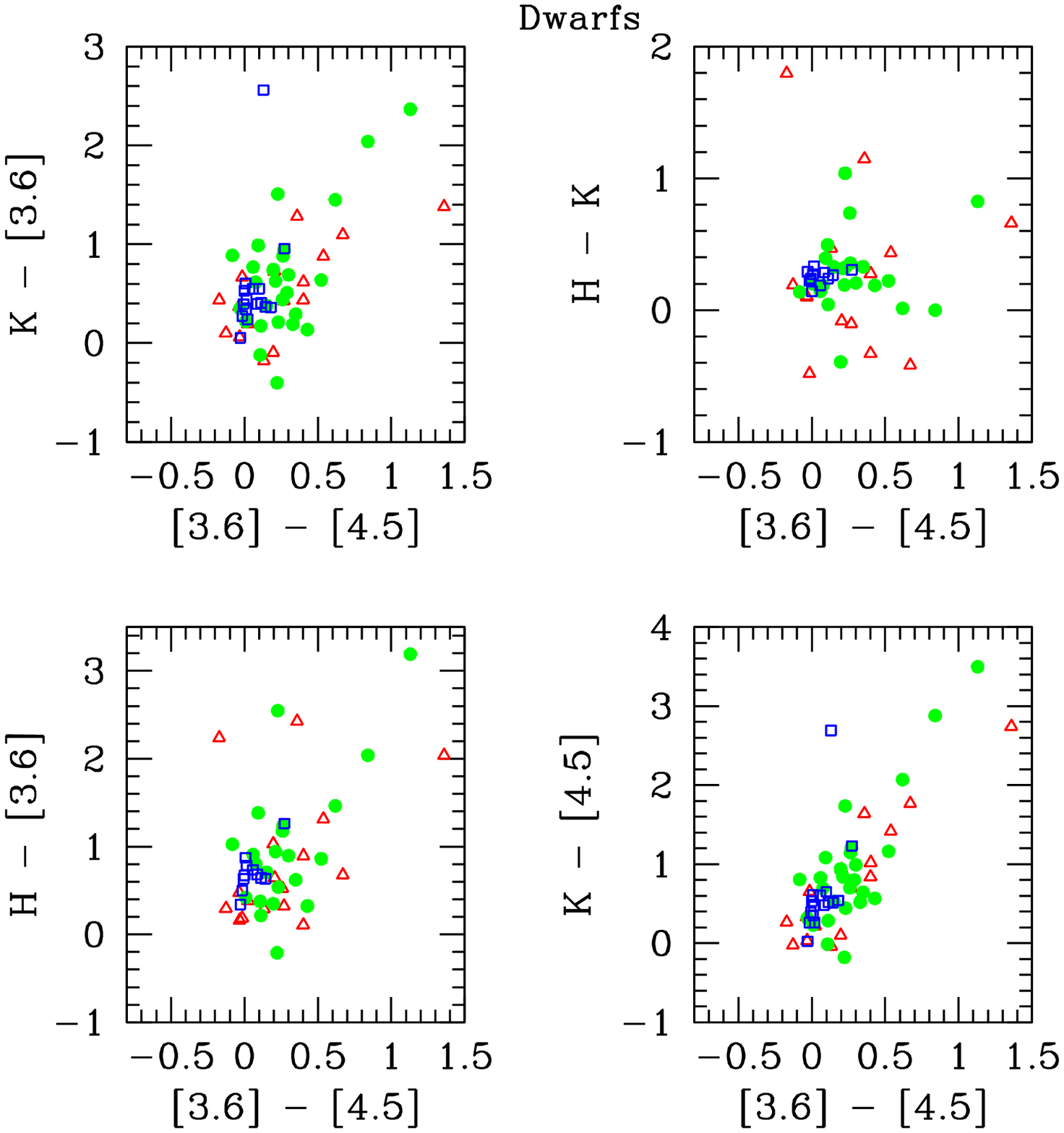}
\includegraphics[width=3.1in]{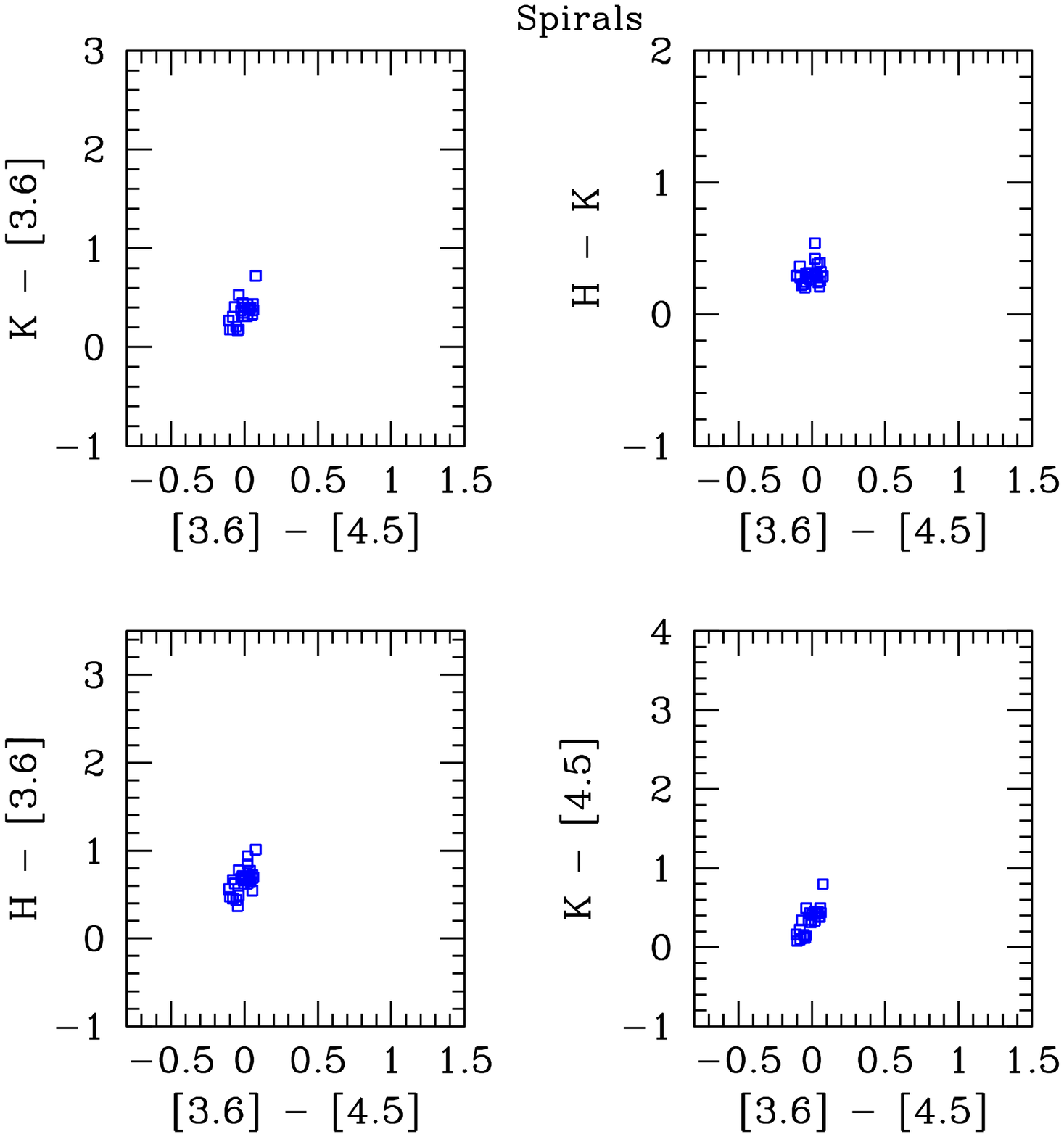}
\caption{Left: Various near-to-mid-infrared color-color plots for our
dwarf
sample galaxies.
The red open triangles are log(O/H) + 12 $<$ 7.8, the green filled circles
7.8 $\le$ log(O/H) + 12 $\le$ 8.2, and the blue open squares
log(O/H) + 12 $>$ 8.2.
Right: Same plots for the spiral sample.
\label{fig12}}
\end{figure*}

\begin{figure*}
\includegraphics[width=3.1in]{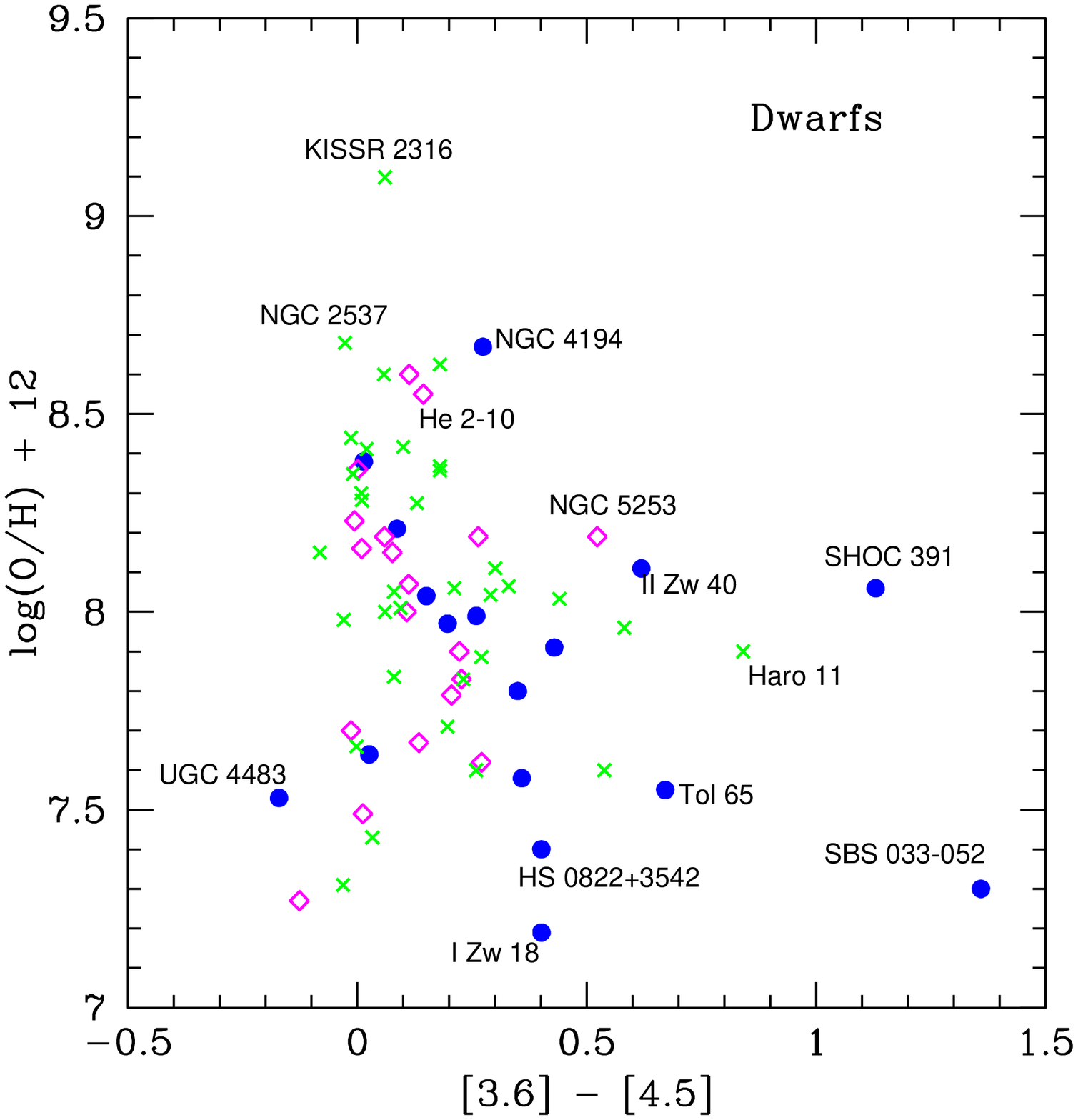}
\includegraphics[width=3.1in]{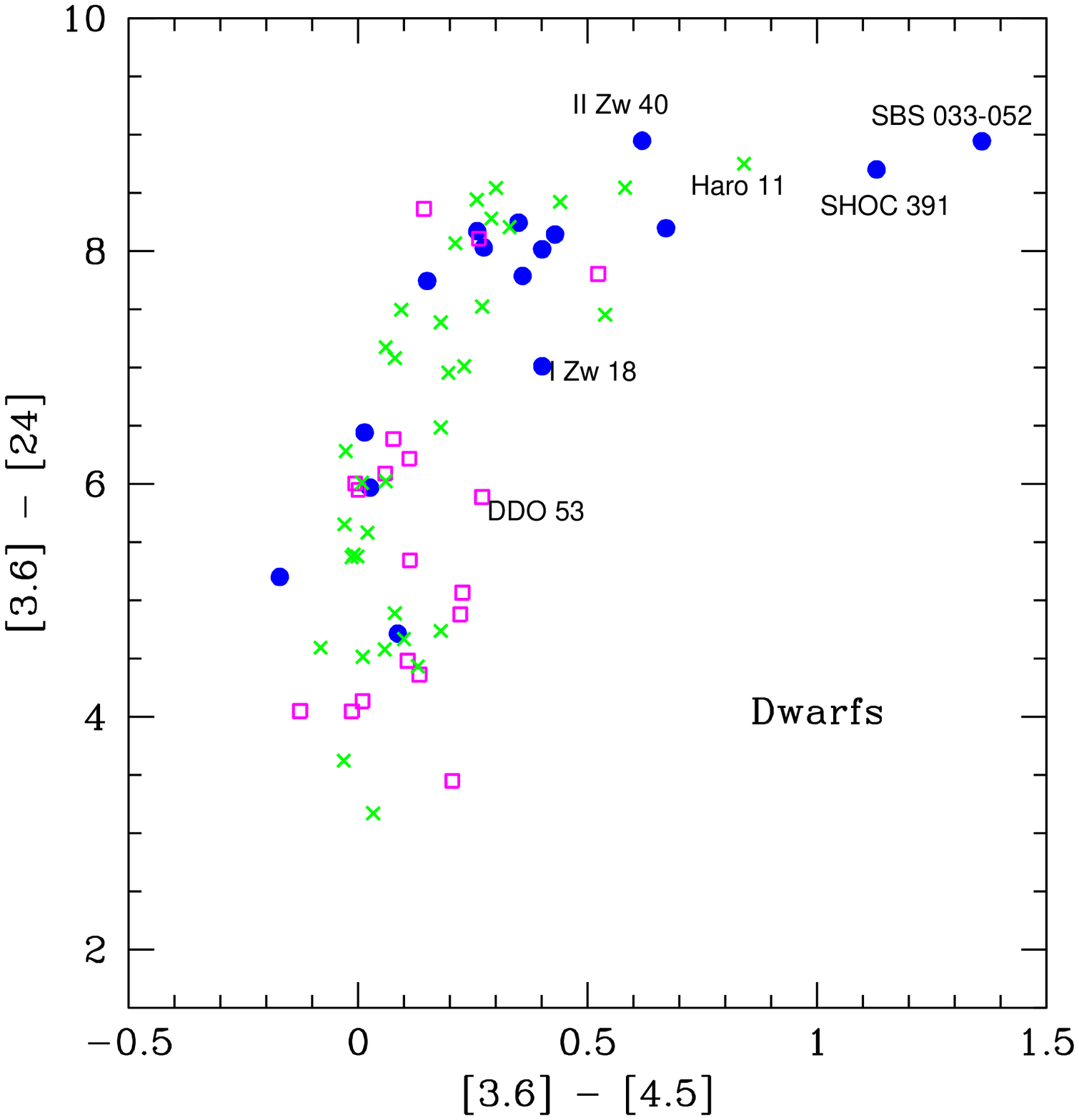}
\caption{Left panel: Same 
as Figure 1a, with the observed [3.6] $-$ [4.5] colors for the 
dwarfs as a function of oxygen abundance, 
but with the data points color-coded
according to morphological types from NED.
The blue filled circles are classified as BCD galaxies or as compact, while the magenta
open diamonds are listed as Irr, Im, I0, IB, or IAB.  The green crosses are galaxies
classified as other types such as HII galaxies or starbursts.
Right panel: Same as Figure 2a, with the [3.6] $-$ [4.5] color
plotted against [3.6] $-$ [24], except with the points
color-coded according to NED morphological type
instead of metallicity.
\label{fig13}}
\end{figure*}

\begin{figure*}
\includegraphics[width=3.1in]{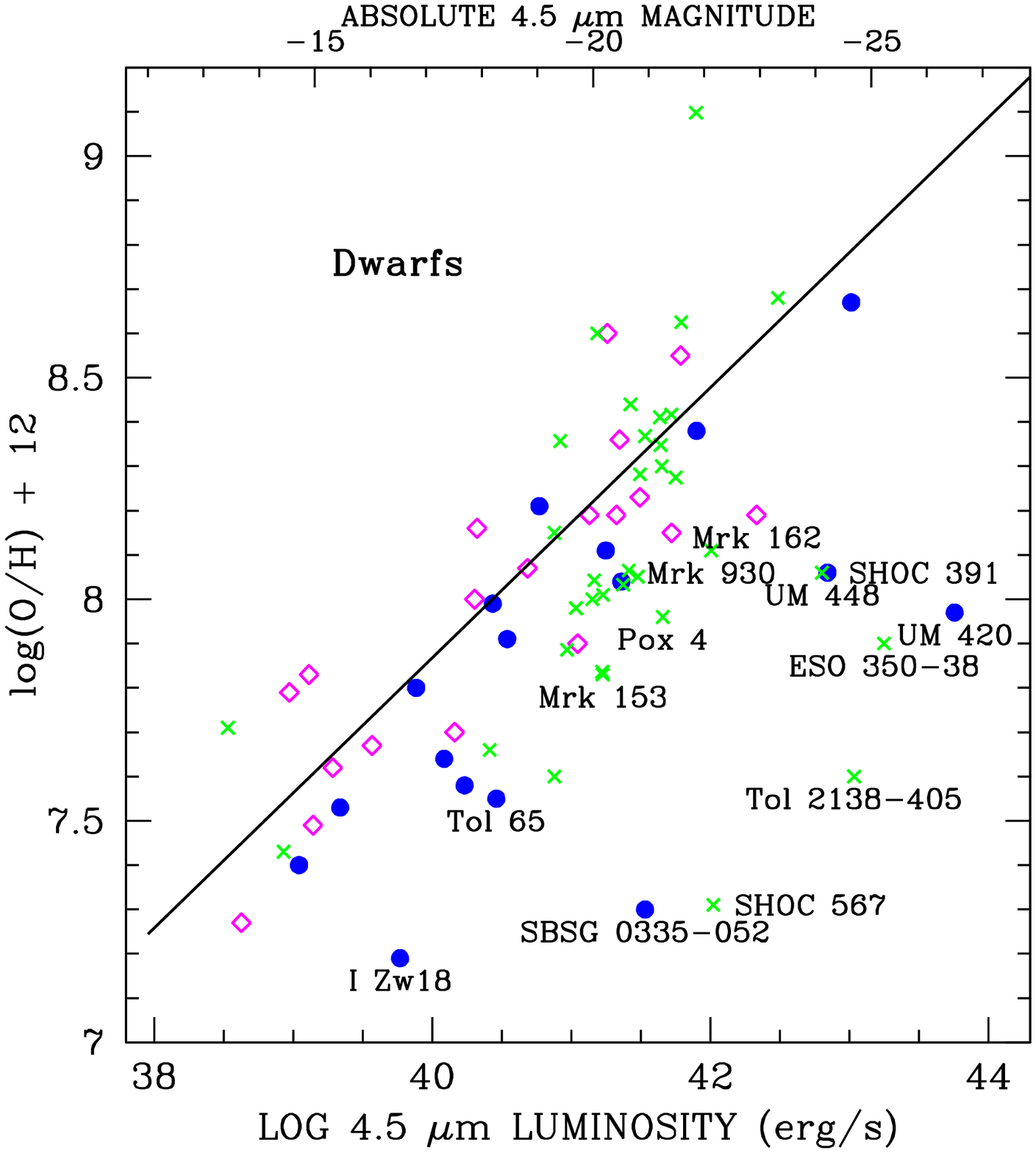}
\includegraphics[width=3.1in]{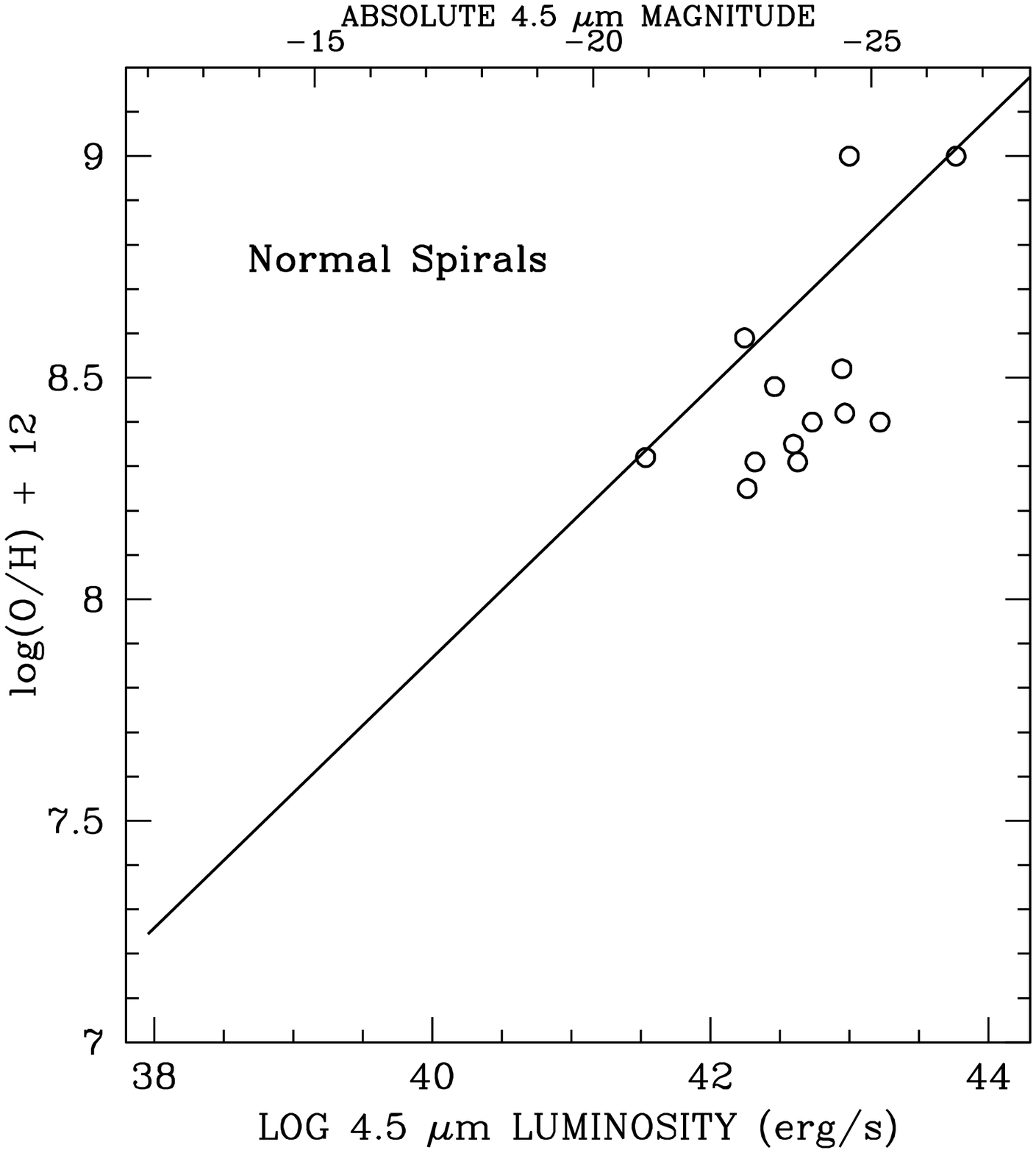}
\caption{
Left panel: the 4.5 $\mu$m luminosity $\nu$L$_{\nu}$ plotted against
oxygen abundance log(O/H) + 12, for our sample dwarfs.
The \citet{lee04} best-fit line for irregular galaxies is shown.
Right panel: a similar figure for our spirals.
The data points are color-coded
according to morphological types from NED as in Figure 13.
The blue filled circles are classified as BCD galaxies or as compact, while the magenta
open diamonds are listed as Irr, Im, I0, IB, or IAB.  The green crosses are galaxies
classified as other types such as HII galaxies or starbursts.
\label{fig14}}
\end{figure*}

\begin{figure*}
\includegraphics[width=3.1in]{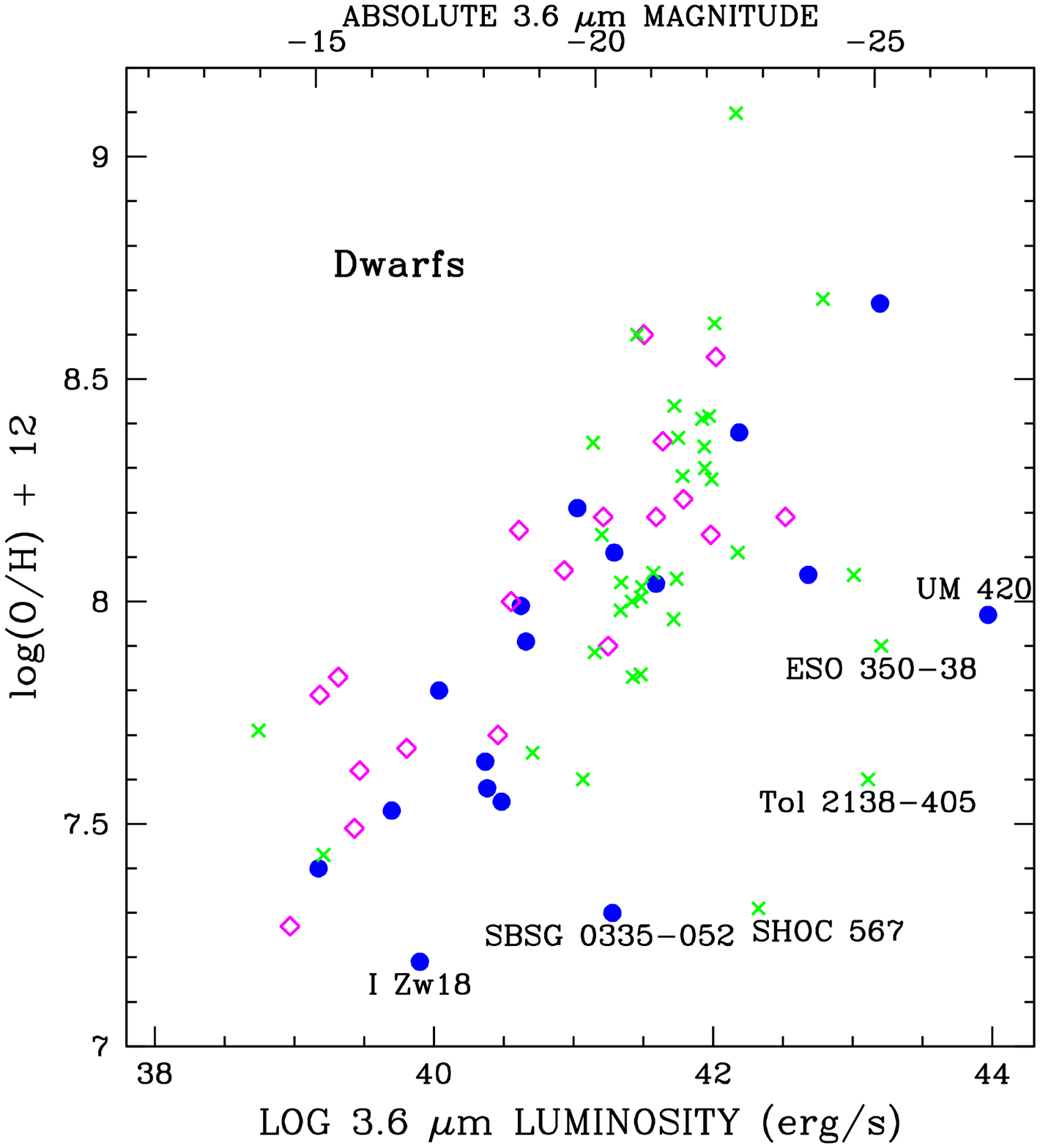}
\includegraphics[width=3.1in]{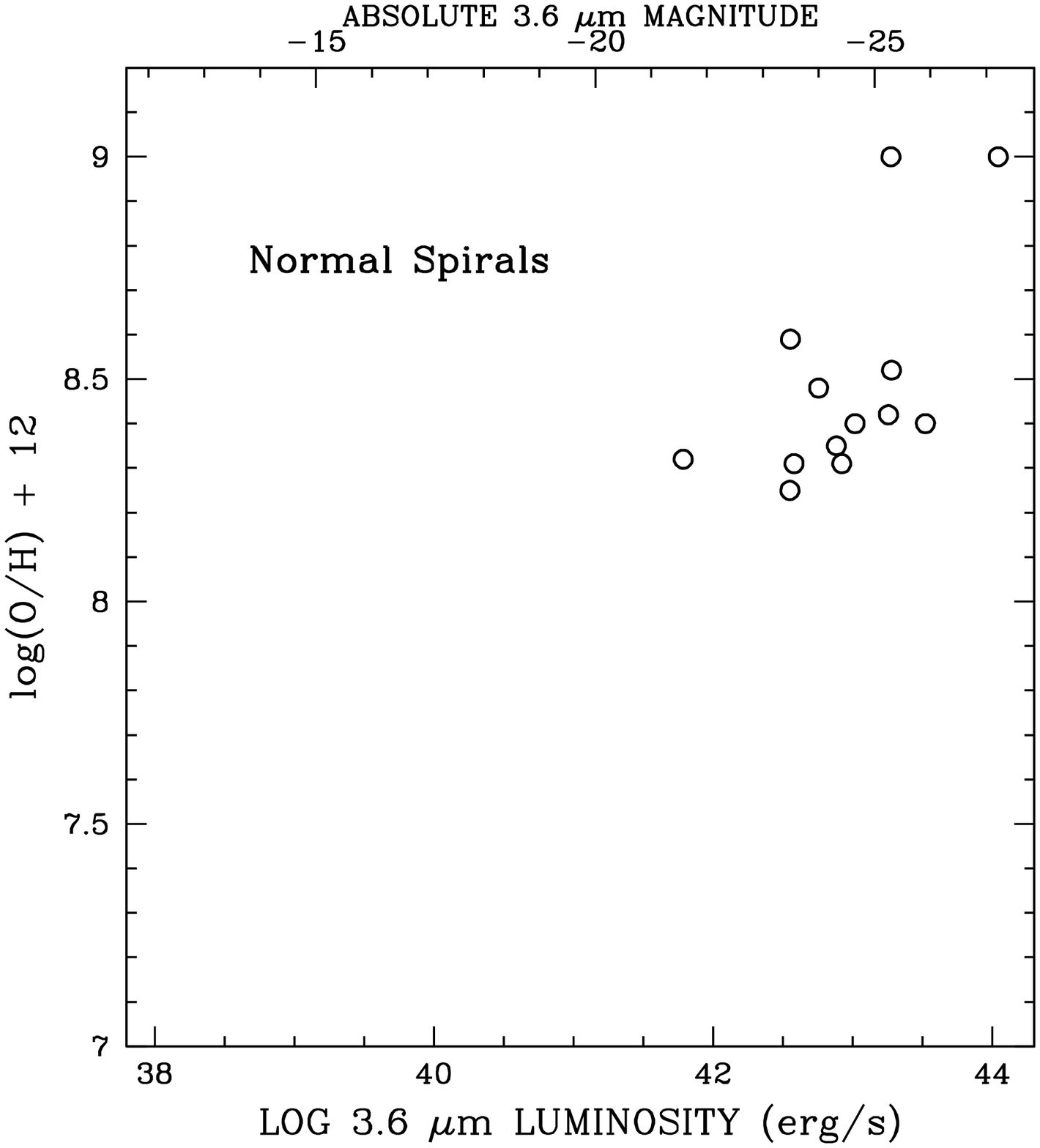}
\caption{
Left panel: the 3.6 $\mu$m luminosity $\nu$L$_{\nu}$ plotted against
oxygen abundance log(O/H) + 12, for our sample dwarfs.
Right panel: a similar figure for our spirals.
The data points are color-coded
according to morphological types from NED as in Figure 13.
The blue filled circles are classified as BCD galaxies or as compact, while the magenta
open diamonds are listed as Irr, Im, I0, IB, or IAB.  The green crosses are galaxies
classified as other types such as HII galaxies or starbursts.
\label{fig15}}
\end{figure*}






\clearpage

\begin{deluxetable}{ccc}
\tabletypesize{\scriptsize}
\tablecaption{Sources of Data used in this Study \label{tbl-1}}
\tablewidth{0pt}
\tablehead{
\colhead{Type of Data} & \colhead{Number} & \colhead{References}\\
& \colhead{Available} & 
}
\startdata
\hline
Spitzer 3.6, 4.5, 5.8, 8.0, and 24 $\mu$m fluxes
&71
&\citet{rosenberg06, rosenberg08, dale07}\\
&&\citet{corbin08,
engelbracht08}\\
Oxygen Abundances&71&\citet{skillman89, lequeux79}\\
&&\citet{masegosa94, miller96}\\
&&\citet{vanZee97, lee04}\\
&&\citet{mendes06, sidoli06}\\
&&\citet{rosenberg06, corbin08}\\
&&\citet{engelbracht08}\\
21 cm HI fluxes&40&\citet{begum06, chengalur06, paturel03}\\
&&\citet{huch05, pustilnik07}\\
H$\alpha$ fluxes&49&\citet{hunter85, hunter94}\\
&&\citet{lehnert95, armus90}\\
&&\citet{gallagher91,
miller94}\\
&&\citet{marlowe97,
mendez99}\\
&&\citet{gil03, james04}\\
&&\citet{pustilnik04, hunter04}\\
&&\citet{begum06, schmitt06}\\
&&\citet{rosenberg06, rosenberg08, 
engelbracht08}\\
H$\alpha$/H$\beta$ ratios&53&\citet{french80, kinman81}\\
&&\citet{campbell86}\\
&&\citet{degrijp92, kennicutt92}\\
&&\citet{vacca92, roennback95}\\
&&\citet{miller96, izotov97, vanZee00}\\
&&\citet{kong02, guseva00}\\
&&\citet{kniazev03, kniazev04, rosenberg06}\\
&&\citet{mous06, izotov07}\\
Br$\alpha$ and/or Br$\gamma$ fluxes&3&\citet{moorwood88, kawara89}\\
&&\citet{dale01, verma03}\\
H$\beta$ equivalent widths&54&\citet{campbell86, bergvall86}\\
&&\citet{terlevich91, kennicutt92, vacca92}\\
&&\citet{roennback95, pustilnik99}\\
&&\citet{guseva00, vanZee00, kong02}\\
&&\citet{kniazev04, salzer05, thuan05}\\
&&\citet{izotov06, mous06}\\
&&\citet{izotov07}\\
9.7 $\mu$m silicate absorption&4&\citet{aitken82, phillips84}\\
&&\citet{thuan99, houck04, brandl06}\\
UBVRI magnitudes&55 &\citet{salzer89, roennback94}\\
&($\ge$3 bands)&\citet{makarova98, makarova99}\\
&&\citet{mendez99b, cairos01, bergvall02}\\
&&\citet{paturel03, gil03}\\
&&\citet{jangren05, noeske03}\\
&&\citet{pustilnik04, mous06}\\
&&\citet{hunter06b}\\
JHK magnitudes&55&\citet{spinoglio95, vanzi00, jarrett03}\\
&($\ge$1 band)&\citet{vaduvescu05,
rosenberg06}\\
&&\citet{mous06,
dale07}\\
&&\citet{engelbracht08}\\
\hline
\enddata




\end{deluxetable}




\end{document}